\documentclass[aps,pre,groupedaddress,amsmath,amssymb,reprint]{revtex4-1}

\usepackage[english]{babel}
\usepackage[utf8]{inputenc}
\usepackage{hyperref}
\usepackage{color}
\usepackage{latexsym}
\usepackage{amsmath}
\usepackage{amssymb}
\usepackage{amsfonts}
\usepackage{bm}

\usepackage{graphicx}
\usepackage{graphics}
\usepackage{caption}
\usepackage{subcaption}
\usepackage{float}
\usepackage{placeins} 
\usepackage{lastpage}
\usepackage[format=plain,justification=raggedright,singlelinecheck=false,font=small,labelfont=bf,labelsep=space]{caption}

\usepackage{array}
\usepackage{xmpmulti}
\usepackage{multirow}

\begin{document}

\title{Effect of the image resolution on the statistical descriptors of heterogeneous media}
\author{René Ledesma-Alonso}
\email{rledesmaalonso@gmail.com}
\affiliation{CONACYT-Universidad de Quintana Roo, Boulevar Bah\'ia s/n, Chetumal, 77019, Quintana Roo, M\'exico}
\author{Romeli Barbosa}
\email{romelix1@gmail.com}
\affiliation{Universidad de Quintana Roo, Boulevar Bah\'ia s/n, Chetumal, 77019, Quintana Roo, M\'exico}
\author{Jaime Ortegon}
\affiliation{Universidad de Quintana Roo, Boulevar Bah\'ia s/n, Chetumal, 77019, Quintana Roo, M\'exico}
\date{\today}

\begin{abstract}
The characterization and reconstruction of heterogeneous materials, such as porous media and electrode materials, involve the application of image processing methods to data acquired by SEM or other microscopy techniques.
Among them, binarization and decimation are critical in order to compute the correlation functions that characterize the micro-structure of the abovementioned materials.
In this study, we present a theoretical analysis of the effects of the image size reduction, due to the progressive and sequential decimation of the original image.
Three different decimation procedures (random, bilinear and bicubic) were implemented and their consequences on the discrete correlation functions (two-points, line-path and pore-size distribution) and the coarseness (derived from the local volume fraction) are reported and analyzed.
The chosen statistical descriptors (correlation functions and coarseness) are typically employed to characterize and reconstruct heterogeneous materials.

A normalization for each of the correlation functions has been performed.
When the loss of statistical information has not been significant for a decimated image, its normalized correlation function is forecast by the trend of the original image (reference function).
In contrast, when the decimated image does not represent the statistical evidence of the original one, the normalized correlation function diverts from the reference function.
Moreover, the equally weighted sum of the average of the squared difference, between the discrete correlation functions of the decimated images and the reference functions, leads to a definition of an overall error.
During the first stages of the gradual decimation, the error remains relatively small and independent of the decimation procedure.
Above a threshold defined by the correlation length of the reference function, the error becomes a function of the number of decimation steps.
At this stage, some statistical information is lost and the error becomes dependent of the decimation procedure. 

These results may help us to restrict the amount of information that one can afford to lose during a decimation process, in order to reduce the computational and memory cost, when one aims to diminish the time consumed by a characterization or reconstruction technique, yet maintaining the statistical quality of the digitized sample.
\end{abstract}
\maketitle

\section{Introduction}

The advancements on image acquisition technology, which direct consequence is the enhanced resolution of scanners, probes and sensors, allows to obtain a huge amount of information and a detailed description of everyday objects.
In addition, the application of this technological progress is not restricted to human scale, since the connection with instruments enables the observation and analysis of objects of scientific interest, in either a much larger scale with a telescope or a much smaller scale with a microscope.
For instance, a scanning electron microscope (SEM) can produce high-resolution images of the microscopic structure of materials, such as porous media, composites and random heterogeneous media (RHM).
Each SEM image, which corresponds to a two-dimensional grayscale map, provides a projection of the microscopic configuration of the phases that constitute the material under examination.
With the employment of the proper techniques~\cite{Torquato2010,Li2014}, this microscopic portrait can be used to characterize the original material and to reconstruct a representative three-dimensional sample that mimics the structure of the heterogeneous media.

3D materials characterization and reconstruction methods based on correlation functions have been broadly accepted since 20 years ago~\cite{Torquato1998a,Torquato1998b,Torquato1999}, and they continue to be used~\cite{Torquato2007,GarmestaniEtal2009,BarbosaEtal2011,BaniassadiEtal2012,HuangLi2013,TahmasebiEtal2012,TahmasebiEtal2013,TahmasebiEtal2015,LiuEtal2013,PantEtal2014}.
Such reconstructions are of great value in a wide variety of fields, mainly for the recovery of the effective properties of the heterogeneous materials~\cite{Torquato,Sahimi,Snarskii}.
The knowledge of these effective properties is required for applications in physics, materials science, engineering, geology and biology, among other unexpected domains such as food industry~\cite{KanitEtal2006} and functional clothing~\cite{Textiles2015}.
A particular case of interest is the characterization, 3D reconstruction and determination of effective properties of the catalytic layer (CL) of a Proton Exchange Membrane Fuel Cell (PEMFC)~\cite{BarbosaEtal2011,MiyamotoEtal2010,ShojaeefardEtal2016,CheVedrine}, which will eventually lead to the design of CLs that maximize the separation, exchange and conduction of chemical species.

Recent studies focus on the determination of the size of the representative volumes to characterize the microstructure of the materials~\cite{HarrisChiuA2015,HarrisChiuB2015} and to simulate and recover the effective properties of heterogeneous materials in different fields, for instance in mechanics~\cite{PelissouEtal2009,MirkhalafEtal2016}.
In previous studies~\cite{Torquato2009}, it has been also found that the size of the observation window $\Omega$, within the domain space $\mathcal{V}$ of the entire two-phase RHM, determines the fluctuation of the local volume fraction $\tau_j$, with respect to the global volume fraction $\phi_j$ of phase $j$.
Special interest has been paid to the determination of the correct size of the image, required to generate an adequate characterization and a representative 3D sample of the material.
Nevertheless, these efforts have been only focused on setting the physical size of the sample, neglecting a fundamental issue: the image resolution, which defines the amount of data and the computing time during the characterization, reconstruction and simulation processes.

High-resolution images, their storage and processing stages require high computing power.
In this context, it is necessary to optimize the amount of information that an image, representative of a material sample, can provide for its satisfactory characterization and 3D reconstruction.
This path is explored in the present study, since we seek to reduce the computational load by decreasing the resolution (number of pixels) of the image, while preserving the statistical information that characterizes the original material sample.
The statistical resemblance between the original image and a reduced-size image is verified by means of statistical descriptors (correlation functions~\cite{Torquato}).

Herein we propose a method based on the correlation length of low-level normalized correlation functions , which indicates the maximum number of steps of a gradual decimation process, that are allowed in order to obtain an image of minimal size (in pixels) without losing statistical information.

This manuscript is organized as follows.
In section~\ref{Sec:Charac} we describe briefly the classic and normalized statistical descriptors and the local volume fraction, which are used to characterize the material samples. 
The generation of original (full-size) images representing heterogenous materials, the gradual decimation process, the corresponding characterization and the computation of the deviation of the reduced-size images with respect to the original image and the coarseness are presented in section~\ref{Sec:Method}.
The implementation of these ideas and procedures on several types of numerically generated binary images of heterogeneous materials has been done and the results are exposed in section~\ref{Sec:Results}.
The main outcome of this manuscript is described in section~\ref{Sec:Algorithm}, an algorithm that performs a statistically sensitive decimation process.
Additionally, an example of a real case application of this methodology, the characterization and decimation of an image representing the catalytic layer of a PEMFC, is presented is section~\ref{Sec:AppSEM}.
Finally, conclusions are given in sec.~\ref{Sec:Conclusions}.

\section{Characterization}
\label{Sec:Charac}

\subsection{Statistical descriptors}

Consider a material with domain $\mathcal{V}\subseteq \mathbb{R}^d$, of volume $V$ and composed by $J=2$ disjoint random phases.
Each phase $j\in\left[0,J-1\right]$ presents a particular domain $\mathcal{V}_j$, a volume fraction $\phi_j$ and an interfacial area per unit volumen $s$.
Therefore, it turns out that
\begin{align}
\bigcup_{j=0}^{J-1}\mathcal{V}_j &=\mathcal{V} \ , &
\bigcap_{j=0}^{J-1}\mathcal{V}_j &=\emptyset \ , &
\sum_{j=0}^{J-1}\phi_j &=1 \ .
\end{align} 

The indicator function $\mathcal{I}_j\left(\bm{x}\right)$ of phase $j$ is defined as
\begin{equation}
\mathcal{I}_j\left(\bm{x}\right)=\begin{cases}
1 & \text{if the point } \bm{x} \text{ lies on } \mathcal{V}_j \\
0 & \text{otherwise}
\end{cases} \ ,
\end{equation}
where $\bm{x}$ is any position in $\mathcal{V}$.
In addition, the set of indicator functions must satisfy the relationship
\begin{equation}
\sum_{j=0}^{J-1}\mathcal{I}_j\left(\bm{x}\right)=1 \ .
\end{equation}

In this work, three low-level correlation functions are used as statistical descriptors: two-point correlation function, line-path correlation function and pore-size distribution function.
The definitions of these statistical descriptors~\cite{Torquato}, which will be recalled in the following, are based on the previously described indicator functions $\mathcal{I}_j$.
Even though these expressions are given for a three-dimensional media, the equivalent formulas for a 2D media are straightforward.

\subsubsection{Two-point correlation function}

The two-point correlation function $S_j^{(2)}(\bm{x}_a,\bm{x}_b)$ is the probability that the two end points  $\bm{x}_a$ and $\bm{x}_b$ of a line segment of length $r$ fall in the same phase $j$.
Considering a statistically homogeneous and isotropic media, this statistical descriptor is no longer a function of the specific positions $\bm{x}_a$ and $\bm{x}_b$, but becomes only a function of the distance $r$, thus being represented by
\begin{align}
S_j^{(2)}\left(r\right) & \equiv  \Big\langle\mathcal{I}_j\left(\bm{x}\right)\mathcal{I}_j\left(\bm{x}+\bm{r}\right)\Big\rangle \ ,
\end{align}
where $\langle\cdot\rangle$ is the expectation value in the domain $\mathcal{V}\subseteq \Re^d$, $\bm{x}$ is any position in $\mathcal{V}$ and $\bm{r}$ is a vector of relative position with magnitude $r$ and an arbitrary direction.
In addition, the absence of long-range order allows us to forecast the following values
\begin{align}
S_j^{(2)}\left(0\right) &=\phi_j \ , & S_j^{(2)}\left(\infty\right) &\rightarrow\phi_j^2 \ .
\end{align}

From this correlation function, we can also define the autocovariance function
\begin{equation}
\chi_j\left(r\right)=S_j^{(2)}\left(r\right)-\phi_j^2 \ ,
\end{equation}
which presents the corresponding values
\begin{align}
\chi_j\left(0\right) &=\phi_j\left[1-\phi_j\right] \ , & \chi_j\left(\infty\right) &\rightarrow 0 \ .
\end{align}

\subsubsection{Line-path correlation function}

The line-path correlation function $L_j\left(\bm{x}_a,\bm{x}_b\right)$ is the probability that a line segment of length $r$ and with end points $\bm{x_a}$ and $\bm{x_b}$ lies wholly in phase $j$.
Considering a statistically homogeneous and isotropic media, this statistical descriptor is no longer a function of the specific positions $\bm{x}_a$ and $\bm{x}_b$, but becomes only a function of the distance $r$, thus reading
\begin{align}
L_j\left(r\right) & \equiv  \left\langle\left\lfloor\int_{0}^{1}\mathcal{I}_j\left(\bm{x}+\alpha\bm{r}\right)d\alpha\right\rfloor\right\rangle
 \ ,
\end{align}
where $\langle\cdot\rangle$ is the expectation value in the domain $\mathcal{V}\subseteq \Re^d$, $\bm{x}$ is any position in $\mathcal{V}$ and $\bm{r}$ is a vector of relative position with magnitude $r$ and an arbitrary direction.
In addition, the absence of long-range order allows us to predict the following values
\begin{align}
L_j\left(0\right) &=\phi_j \ , & L_j\left(\infty\right) &\rightarrow 0 \ .
\end{align}

\subsubsection{Pore-size probability density function}

The pore-size probability density function $P_j\left(r\right)$ is the likelihood of finding the nearest point on the $ji$ interface (with $i\neq j$) at a distance $r$ from a point $\bm{x}$ in phase $j$.
This statistical descriptor is described by the expression
\begin{align}
P_j\left(r\right) &\equiv
\left\langle\left\lfloor\dfrac{1}{V_D}\int_{V_D}\mathcal{I}_j\left(\bm{x}+\alpha\bm{r}\right)\, dV_D\right\rfloor\right\rangle\, \ ,
\end{align}
where $V_D$ and $dV_D$ are the volume of a $D$-dimensional unit sphere and a differential volume, given either by $dV_2=\alpha d\alpha d\theta$ in a local $(D=2)$-dimensional domain, or by $dV_3=\alpha^2\sin\left(\theta\right)d\alpha d\theta d\varphi$ in a local $(D=3)$-dimensional domain, in both cases considering that $\alpha\in\left[0,1\right]$, $\theta\in\left[0,2\pi\right]$ and $\varphi\in\left[0,\pi\right]$.
Particularly in this expression, $\langle\cdot\rangle$ is the expectation value in the domain $\mathcal{V}_j$, $\bm{x}$ is any position in $\mathcal{V}_j$ and $\bm{r}$ is a vector of relative position with magnitude $r$ and an arbitrary direction.

In addition, the absence of long-range order allows us to estimate the following values
\begin{align}
P_j\left(0\right) &=\dfrac{s}{\phi_j} \ , & P_j\left(\infty\right) &\rightarrow 0 \ .
\end{align}

\subsection{Normalized statistical descriptors}
\label{subsec:Nsd}

From the previously mentioned statistical descriptors, the normalized statistical descriptors $F_{\beta,j}$, which are functions of the distance $r$, are defined as 
\begin{subequations}
\begin{align}
F_{1,j}\left(r\right) &=\dfrac{\chi_j\left(r\right)}{\phi_j\left(1-\phi_j\right)}=\chi_j^{\ast}\left(r\right) \ , \\
F_{2,j}\left(r\right) &=\dfrac{L_j\left(r\right)}{\phi_j}=L_j^{\ast}\left(r\right) \ , \\
F_{3,j}\left(r\right) &=\dfrac{\phi_j P_j\left(r\right)}{s}=P_j^{\ast}\left(r\right) \ .
\end{align}
\label{Eq:NCF}
\end{subequations}
The value of the index $\beta$ is respectively associated with the type of descriptor: the autocovariance function for $\beta=1$, the normalized line-path correlation function for $\beta=2$ and the normalized pore-size distribution function for $\beta=3$.

\subsection{Local volume fraction}

While the volume fraction $\phi_j$ of phase $j$ is considered to be a global quantity, at a local level the volume fraction of phase $j$, represented by $\tau_j\left(\bm{x},l\right)$, depends on the position $\bm{x}$ and size $l$ of a given observation window $\Omega_l$.
Considering an observation window defined by 
\begin{equation}
\psi\left(\bm{x},l\right)=\begin{cases}
1 \ , & \text{for } \bm{x} \in \Omega_l \\
0 \ , & otherwise \ ,
\label{Eq:ObWin}
\end{cases}
\end{equation}
the local volume fraction reads
\begin{equation}
\tau_j\left(\bm{x},l\right)=\dfrac{1}{V_l}\int\mathcal{I}_j\left(\bm{z}\right) \, \psi\left(\bm{z-\bm{x}},l\right) \, d\bm{z} \ ,
\end{equation}
where $V_l=V_l\left(l\right)$ is the volume of the observation window.
Departing from these definitions, the variance of the local volume fraction is recovered with
\begin{equation}
\sigma^2_j\left(l\right)=\left\langle \tau^2_j\left(\bm{x},l\right) \right\rangle-\phi_j^2 \ ,
\label{Eq:varLVF}
\end{equation}
where $\langle\cdot\rangle$ denotes the average among different positions of the observation window within the global domain.
Finally, the coarseness $C_j$, which is the normalized standard deviation from the global volume fraction of phase $j$ for a given size $l$ of the observation window $\Omega_l$, is defined as
\begin{equation}
C_j\left(l\right)=\dfrac{\sigma_j\left(l\right)}{\phi_j} \ .
\end{equation}
This quantity gives a measure of the local fluctuations of the volume fraction~\cite{Torquato2009}.

For the particular case of two phases media, we find the following relations
\begin{align}
\sigma\left(l\right) &=\sigma_0\left(l\right)=\sigma_1\left(l\right) \ , & C_0\left(l\right) &=\left[\dfrac{\phi_1}{\phi_0}\right]C_1\left(l\right)\ .
\end{align}
In addition, an infinitesimally small window and an infinitely large window yield
\begin{align}
C_1\left(0\right) &=\dfrac{\sqrt{\phi_0\phi_1}}{\phi_1} \ , & C_1\left(\infty\right) &\rightarrow0 \ ,
\end{align}
respectively.
Finally, considering this properties, a normalized coarseness for a two-phase material is defined as follows
\begin{equation}
C^{\ast}\left(l\right)=\dfrac{C_j\left(l\right)}{C_j\left(0\right)}=\dfrac{\sigma\left(l\right)}{\sqrt{\phi_0\phi_1}} \ .
\end{equation}

\section{Methodology}
\label{Sec:Method}

Consider two phases planar materials formed by a void phase ($j=0$) and a solid phase ($j=1$).
A material will be represented by a 2D binary image, a matrix $\mathbf{A}$ with $M$ rows and $N$ columns, representing a discretized binary field.
Thus, each element or pixel $A\left(m,n\right)$, with $m,n\in\mathbb{Z}^{+}$ in the ranges $m\in\left[0,M-1\right]$ and $n\in\left[0,N-1\right]$, presents a value of either 0 or 1.

First, a realization $w$ of the possible states of the material, a configuration with domains $\mathcal{V}_j\left(w\right)$, is generated.
The result is an original or full-resolution image of the material, represented by the matrix $\mathbf{A}_0=\mathbf{A}_0\left(w\right)$, with $w\in\left[1,W\right]$ and $W=20$ being the total number of random configurations of the same material.
From $\mathbf{A}_0$, the normalized statistical descriptors $F_{\beta,j,0}$ are computed and taken as the reference functions of the realization $w$.
Afterwards, the full-resolution image $\mathbf{A}_0$ is subjected to a progressive and sequential decimation process.
At the $k$th decimation stage, with $k\in\left[1,8\right]$, the resulting matrix $\mathbf{A}_k$ is recovered and the corresponding normalized statistical descriptors $F_{\beta,j,k}$ are determined.
A comparison of $F_{\beta,j,k}$ with the reference $F_{\beta,j,0}$ leads to estimate the deviation of the specific descriptor $\beta$ and the phase $j$, at the $k$-th decimation stage.
As well, the overall error of the decimated image $\mathbf{A}_k$, at the $k$th stage, with respect to the full-resolution image $\mathbf{A}_0$ is obtained by taking the addition of the deviations of all the descriptors $\beta=1,2,3$ and all the phases $j=0,1$.

In this section, details about the generation of full-resolution images, the decimation process and the calculation of the descriptors deviations and the overall error are presented.  

\subsection{Full-resolution images}

\begin{figure}
\centering
\includegraphics[width=0.5\textwidth]{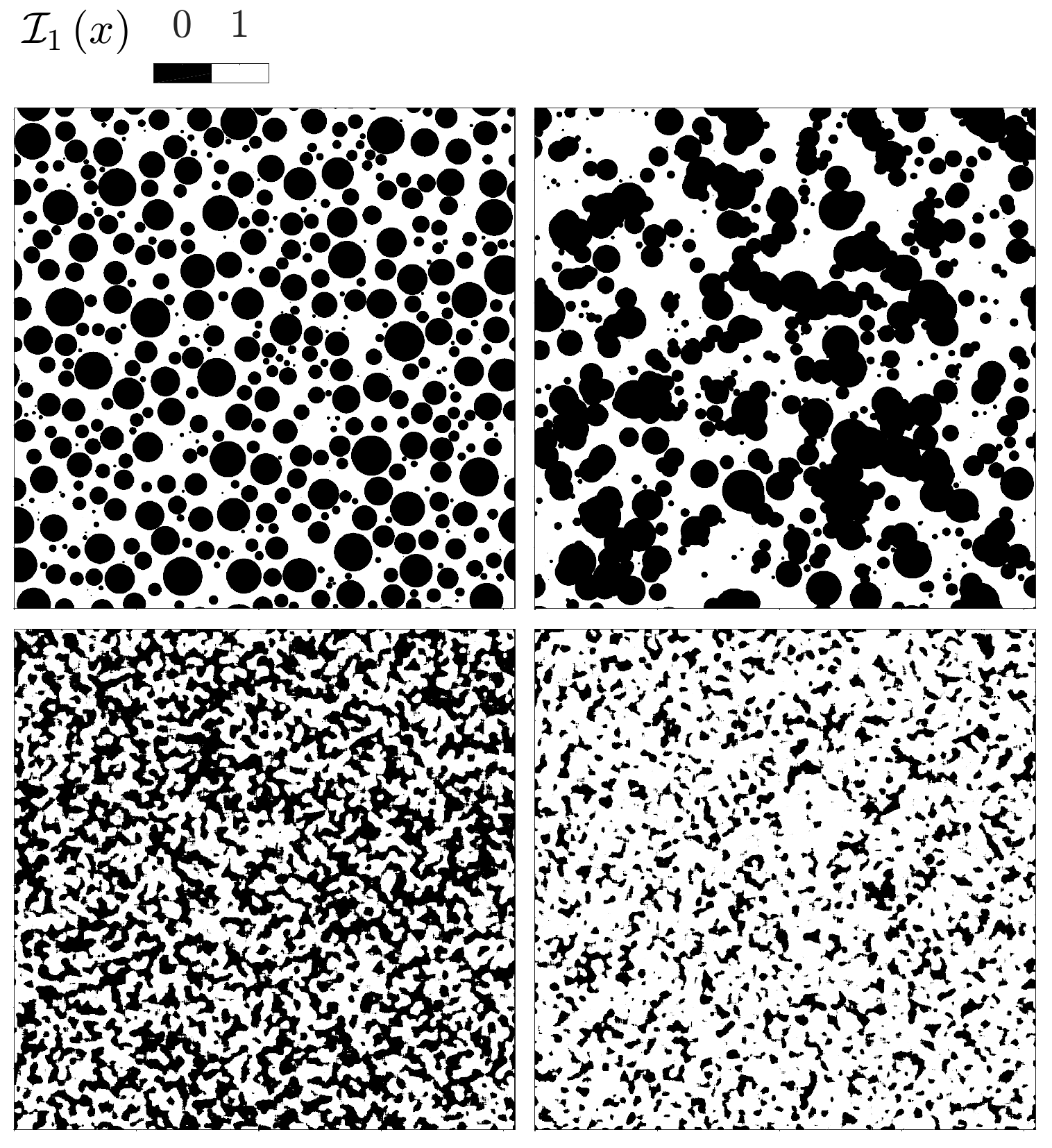}
\caption{Statistically isotropic heterogenous materials: impenetrable disks (ID - top left), partially overlapping disks (OD - top right) and coarsed grained random heterogenous materials obtained with a Laplacian of Gaussian kernel (LoGK1 - bottom left and LoGK3 - bottom right), for two different values of the level-cut threshold. The corresponding surface fractions, as well as the parameters used for their conformation, are given in Tables~\ref{Tab:Disks} and \ref{Tab:CoarseGrained}.}
\label{Fig:Materials}
\end{figure}

The full-resolution images were constructed following two processes leading to different types of images.
An example of each type of the generated images is shown in Figure~\ref{Fig:Materials}.
The first process involves the positioning of disks in two variants: 1a) impenetrable disks, and 1b) partially overlapping disks.
The second produces random heterogenous materials (RHM) by coarse graining a matrix of randomly generated values.
Different sets of parameters for the coarse graining were employed to obtain RHMs with different values of the surface fractions $\phi_j$. 
All these processes are described in the following.

\subsubsection{Impenetrable and partially overlapping disks}

\begin{table}
\caption{Average specifications of the two phases planar materials generated with impenetrable (ID) and partially overlapping (OD) disks, using $r_{max}=250$, $r_{min}=1$, $\mu=50$ and $\sigma=60$ for the radii distribution given by eq.~\eqref{Eq:FNdist}.}
\begin{ruledtabular}
\renewcommand{\arraystretch}{1.2}
\begin{tabular}{ccccc}
& \# of disks & \multicolumn{3}{c}{Surface fraction} \\
 \cline{3-5}
Type & $I$ & $\langle\phi_0\rangle$ & $\langle\phi_1\rangle$ & $\sigma_{\phi}$ \\
\hline
ID & 491 & 0.4999 & 0.5001 & $7.86\times10^{-6}$ \\
OD & 640 & 0.5059 & 0.4941 & $6.11\times10^{-3}$
\end{tabular}
\end{ruledtabular}
\label{Tab:Disks}
\end{table}

A number $I$ of disks were placed, which radii have been chosen following a discrete version of the Folded-Normal distribution.
The proposed discrete distribution is described by
\begin{align}
\mathcal{P}_r\left\{r_t\right\} &=\dfrac{1}{I}\left\lfloor I\, \sqrt{\frac{2}{\pi\sigma^2}}\, \exp\left(-\dfrac{r_t^2+\mu^2}{2\sigma^2}\right)\, \cosh\left(\dfrac{r_t \mu}{\sigma^2}\right)\right\rceil \ ,
\label{Eq:FNdist}
\end{align}
where $\lfloor\cdot\rceil$ indicates the application of the nearest integer function, $\mu$ and $\sigma$ are the location and scale parameters, and $r_t =r_{min}+t$ is the discrete radius of bin $t$, which range is $t\in\left[0,r_{max}-r_{min}\right]$ and with $r_{max}$ and $r_{min}$ being the maximum and minimum radii, respectively.
Therefore, the radius $r_i$ of the $i$th disk is randomly chosen to be equal to some radius $r_t$.

The space region $\mathbb{R}^2$ in which the disks can be placed has been chosen to be a square, of area $L^2$, within the ranges $x\in\left[0,L\right]$ and $y\in\left[0,L\right]$.
For the $i$th disk, its center coordinates $\left(x_i,y_i\right)$ have been selected randomly within the ranges $x_i\in\left[0,L\right]$ and $y_i\in\left[0,L\right]$, surveying that the distance between its center and that of the previously placed disks, given by
\begin{align}
d &=\sqrt{\left(x_h-x_i\right)^2+\left(y_h-y_i\right)^2}\ , &\text{ for } 1\leq h<i \ ,
\end{align}
remains with a value
\begin{subequations}
\begin{align}
d &\geq\left(r_h+r_i\right) \ , &\text{ for impenetrable disks} \ , \\
d &>\left|r_h-r_i\right|\ , &\text{for partially overlapping disks} \ .
\end{align}
\end{subequations}

The matrix $\mathbf{A}_0$ with equal number of rows and columns, $M_0=N_0=4096$ is created with pixel values $A_0\left(m,n\right)=0$ for all $m\in\left[0,M_0-1\right]$ and $n\in\left[0,N_0-1\right]$.
Now, considering the coordinates of each pixel $A_0\left(m,n\right)$, being given by $x_{mn}=n+1/2$ and $y_{mn}=m+1/2$, the value of the pixel, which is initially assigned to be $A_0\left(m,n\right)=0$, may change according to the following statement:
if  $\sqrt{\left(x_{mn}-x_i\right)^2+\left(y_{mn}-y_i\right)^2}\leq r_i$ then $A_0\left(m,n\right)=1$; otherwise $A_0\left(m,n\right)$ keeps its previous value.
Additionally, periodic boundary conditions were considered in both $x$ and $y$ directions.
The process is repeated for all the disks, from $i=0$ to $i=I-1$, and the binary matrix $\mathbf{A}_0$ is obtained.

Following abovementioned process, a total of $W=20$ random configurations of impenetrable disks (ID) and partially overlapping disks (OD), which specifications are presented in table~\ref{Tab:Disks}, have been gathered.
The number of disks $I$ and average surface fractions $\langle\phi_0\rangle$ and $\langle\phi_1\rangle$ for phase $j=0$ and $j=1$, respectively, and their standard deviation $\sigma_{\phi}$ are given in the abovementioned table.

\subsubsection{Random heterogeneous material}

\begin{table}
\caption{Average specifications of the two phases planar materials generated by coarse graining with a Laplacian of Gaussian kernel (LoGK) of size $a=75$, specified in eqs.~\eqref{Eq:LK}.}
\begin{ruledtabular}
\renewcommand{\arraystretch}{1.2}
\begin{tabular}{ccccc}
LoGK & Threshold & \multicolumn{3}{c}{Surface fraction} \\
 \cline{3-5}
\# & $a_0$ & $\langle\phi_0\rangle$ & $\langle\phi_1\rangle$ & $\sigma_{\phi}$ \\
\hline
1 & 127.50 & 0.4969 & 0.5031 & $6.81\times10^{-3}$ \\
2 & 127.00 & 0.3272 & 0.6728 & $5.63\times10^{-3}$ \\
3 & 126.75 & 0.2475 & 0.7525 & $6.14\times10^{-3}$
\end{tabular}
\end{ruledtabular}
\label{Tab:CoarseGrained}
\end{table}

A matrix $\mathbf{A}_0^{\ast}$ is created with an equal number of rows and columns, $M_0=N_0=4096$.
An intensity, given by the discrete random variable $a\in\left[0,255\right]$ with a uniform probability function $\mathcal{P}_a\left\{a\right\}=1/256$, is assigned to each pixel $A_0^{\ast}\left(m,n\right)$, with $m\in\left[0,M_0-1\right]$ and $n\in\left[0,N_0-1\right]$.
A blurred matrix $\mathbf{A}_0^{\Box}$ is obtained by performing the convolution of the matrix $\mathbf{A}_0^{\ast}$ and a square-shaped kernel $\boldsymbol{\mathcal{K}}$:
\begin{equation}
A_0^{\Box}\left(m,n\right)=\displaystyle\sum_h \displaystyle\sum_i \mathcal{K}\left(h,i,b\right)A_0^{\ast}\left(m+h,n+i\right)
\end{equation}
where $b$ is the size of the kernel, an odd positive integer $\mathbb{Z}^{+}$, which yields the ranges $h\in\left[-\lfloor b/2\rfloor,\lfloor b/2\rfloor\right]$ and $i\in\left[-\lfloor b/2\rfloor,\lfloor b/2\rfloor\right]$.
Herein, the discrete Laplacian of Gaussian kernel (LoGK), described by
\begin{subequations}
\begin{align}
\mathcal{L}\left(h,i,b\right) &=\left(1-\dfrac{h^2+i^2}{2\left[b/2\right]^2}\right)\exp\left(-\dfrac{h^2+i^2}{2\left[b/2\right]^2}\right) \ , \\
\mathcal{K}\left(h,i,b\right) &=\dfrac{\mathcal{L}\left(h,i,b\right)}{\displaystyle\sum_h \displaystyle\sum_i \mathcal{L}\left(h,i,b\right)} \ ,
\end{align}
\label{Eq:LK}
\end{subequations}
has been employed.
Finally, a binary matrix $\mathbf{A}_0$ is obtained by level-cutting the matrix $\mathbf{A}_0^{\Box}$, using a threshold value $a_0$ as follows
\begin{equation}
A_0\left(m,n\right)=\begin{cases}
0 & \text{ if } A_0^{\Box}\left(m,n\right)<a_0 \ , \\
1 & \text{ if } A_0^{\Box}\left(m,n\right)\geq a_0 \ .
\end{cases}
\end{equation}
Both, the threshold value, which is restricted to the range $a_0\in\left[0.0,255.0\right]$, and the kernel size $b$ are directly responsible for the structural characteristics of the binary matrix $\mathbf{A}_0$.

Following abovementioned process, two sets of parameters were proposed to generate two different types of LoGK coarsed grained RHMs, both with $b=75$ but differing in their value of $a_0$.
A total of $W=20$ random configurations of each LoGK material, which specifications are presented in table~\ref{Tab:CoarseGrained}, have been gathered.
The kernel size $b$, threshold value $a_0$ and average surface fractions $\langle\phi_0\rangle$ and $\langle\phi_1\rangle$ for phase $j=0$ and $j=1$, respectively, and their standard deviation $\sigma_{\phi}$ are presented.

\subsection{Decimation}

\begin{figure*}
\centering
\includegraphics[width=0.88\textwidth]{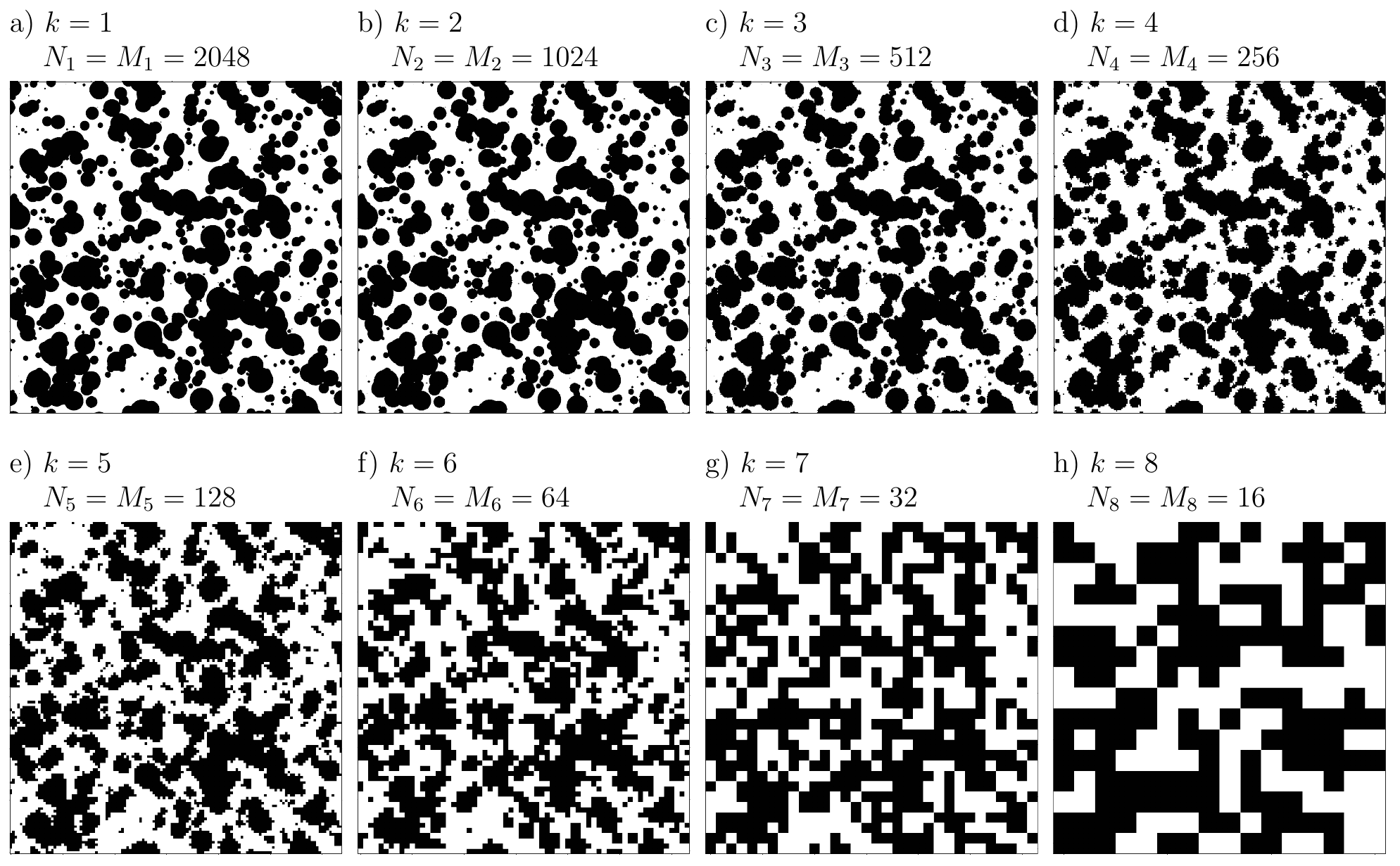}
\caption{Consecutive stages [from a) $k=1$ to h) $k=8$] following a random decimation process, departing from the full resolution image ($k=0$ with $N_0=M_0=4096$) of partially overlapping disks presented in Figure~\ref{Fig:Materials}.}
\label{Fig:Decimation}
\end{figure*}

As it has been previously mentioned, a progressive and sequential decimation process has been implemented, starting from the step $k=1$ and increasing $k$ by one integer, until the final step $K$ is performed.
At each $k$th decimation step, the matrix $\mathbf{A}_{k-1}$, with $M_{k-1}$ rows and $N_{k-1}$ columns, evolves towards a new matrix $\mathbf{A}_k$, where the number of rows and columns is reduced to $M_k=M_{k-1}/2$ and $N_k=N_{k-1}/2$.
In other words, each $k$-decimation step provokes a reduction of the number of rows and columns to $M_k=M_0/\left(2^k\right)$ and $N_k=N_0/\left(2^k\right)$, respectively.
In the present work, defining $K=8$ and taking $M_0=N_0=4096$ leads to images at the final decimation step of dimensions $M_K=N_K=16$.

Each decimation process is performed by the application of one of the three methods: random, bilinear or bicubic decimation~\cite{Bovik,Gonzalez}.
An example of the decimation process, in particular with the application of the random method, is shown in Figure~\ref{Fig:Decimation}.
The three decimation methods are briefly explained in the following.



\subsubsection{Random}

The element in the $m$th row and $n$th column of the matrix $\mathbf{A}_k$ is obtained by performing the $k$th decimation step expressed by
\begin{subequations}
\begin{equation}
A_k\left(m,n\right) = A_{k-1}\left(2m+\xi_m,2n+\xi_n\right) \ ,
\end{equation}
with
\begin{align}
\xi_m &=\lfloor\xi/2\rfloor \ , &
\xi_n &=\xi-2\lfloor\xi/2\rfloor \ ,
\end{align}
\end{subequations}
where $\lfloor\cdot\rfloor$ indicates the application of the floor function, $\xi$ is a discrete random variable in the range $\xi\in\left[0,3\right]$, with a discrete uniform distribution $\mathcal{P}_{\xi}\left\{\xi\right\}=1/4$.

\subsubsection{Bilinear}

The element in the $m$th row and $n$th column of the matrix $\mathbf{A}_k$ is the resulting value from a four elements average, followed by a rounding operation, described by the $k$th decimation step formula
\begin{subequations}
\begin{equation}
A_k\left(m,n\right) = \begin{cases}
0 & \text{if } \alpha_{mn}<1/2 \\
1 & \text{if } \alpha_{mn}\geq1/2 \\
\end{cases} \ ,
\end{equation}
with
\begin{equation}
\alpha_{mn} =\frac{1}{4}\sum_{i=0}^1\sum_{j=0}^1 A_{k-1}\left(2m+i,2m+j\right) \ .
\end{equation}
\end{subequations}

\subsubsection{Bicubic}

The element in the $m$th row and $n$th column of the matrix $\mathbf{A}_k$ is the outcome of the bicubic interpolation, performed in the $k$th decimation step, described by the following expressions
\begin{subequations}
\begin{align}
A_k\left(m,n\right) &= \begin{cases}
0 & \text{if } \alpha_{mn}<1/2 \\
1 & \text{if } \alpha_{mn}\geq1/2 \\
\end{cases} \ ,
\end{align}
with
\begin{align}
\alpha_{mn} &=G\left(q_0,q_1,q_2,q_3\right) \ ,
\end{align}
where, the function $G$ is the linear combination of its arguments given by 
\begin{equation}
G\left(g_0,g_1,g_2,g_3\right)=\dfrac{1}{16}\left[-g_0+9g_1+9g_2-g_3\right]\ .
\end{equation}
\end{subequations}
Herein, $q_0=G(A_{k-1})$ is the value obtained from the application of the function $G$ to the four elements of $A_{k-1}$ in the columns from $2n-1$ to $2n+2$ and in the $\left(2m-1\right)$th row.
In turn, $q_1=G(A_{k-1})$, $q_2=G(A_{k-1})$ and $q_3=G(A_{k-1})$ are obtained equivalently from the same columns of $A_{k-1}$, but for the $\left(2m\right)$th, $\left(2m+1\right)$th and $\left(2m+2\right)$th rows, respectively.
The previously formulas are valid to obtain the values of $\left\{q_0,q_1,q_2,q_3\right\}$ when $m\in\left[1,M_{k-1}-2\right]$ and $n\in\left[1,N_{k-1}-2\right]$.
During the bicubic decimation procedure, particular attention has to be made for the borders of the image, for which $m=0$, $m=M_{k-1}-1$, $n=0$ or $n=N_{k-1}-1$, as it is explained in detail elsewhere (see Supplementary Material).

\subsection{Discrete normalized statistical descriptors}

\begin{figure*}
\centering
\includegraphics[width=0.95\textwidth]{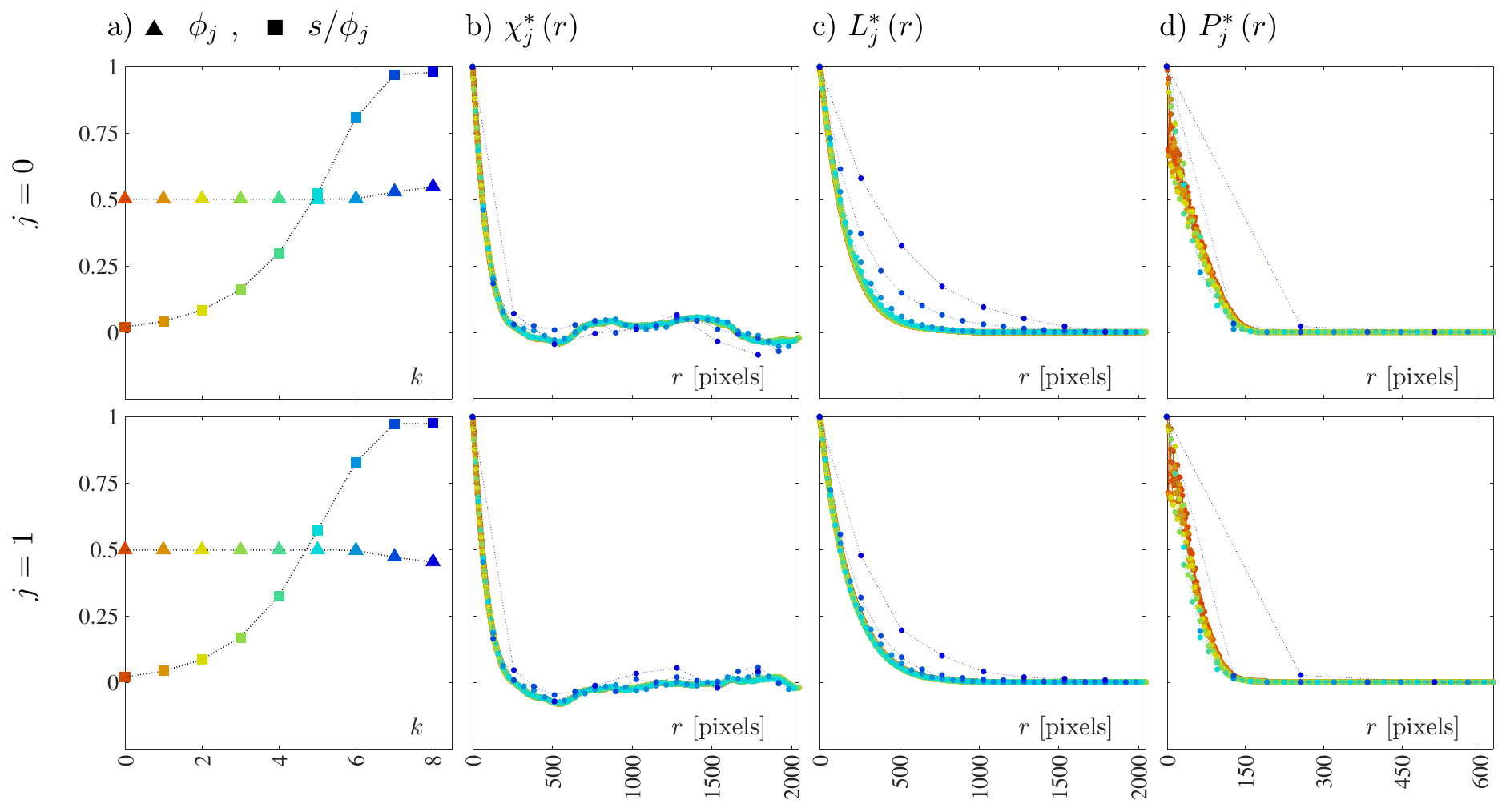}
\caption{Statistical descriptors along the random decimation process presented in Figure~\ref{Fig:Decimation}. Top and bottom rows represent the descriptors for phases $j=0$ and $j=1$ respectively, whereas each column indicates the type of descriptor: a) surface fraction $\phi_j$ and specific interface area per phase surface fraction $s/\phi_j$ as functions of the $k$-th decimation step, b) normalized autocovariance function $F_{1,j}=\chi_j$, c) normalized line-path correlation function $F_{2,j}=L_j$ and d) normalized pore-size distribution function $F_{3,j}=P_j$. The statistical descriptors in columns b)-d) are presented as functions of the distance $r$. Each color of the points in column a) and the curves in columns b)-d) correspond to a different $k$-th decimation step, given by the horizontal position of the point in column a).}
\label{Fig:Descriptors}
\end{figure*}

For all the types of materials, disks and RHMs, the computation of the statistical descriptors $\left\{\chi_j,L_j,P_j\right\}$ of all the images, full-resolution $\mathcal{A}_0$ and decimated $\mathcal{A}_k$, was performed following a well-know method proposed for isotropic digitized systems~\cite{Torquato1998a,Torquato1998b}.
Under the prescribed procedure, the distance is measured in units of pixels, and thus the retrieved distance data is represented by the discrete variable $r_l$, that contains only integral values, with the range $l\in\left[1,N_k^{\dagger}\right]$ and considering the limit $N_k^{\dagger}\approx N_k/2$ to avoid the effects of periodic boundaries or finite size domains.
Therefore, the pixel size of each image must be taken into account in order to recover $r$, an invariant distance variable, from $r_{k,l}$, the distance variable associated with the $k$th decimation step.
For the decimated image $\mathcal{A}_k$, the pixel size is given by $\mathcal{L}_k=N_0/N_k$, whereas the reference scale $\mathcal{L}_0=1$ is the pixel size in the full-resolution image $\mathcal{A}_0$.
Consequently, one finds that $r=\mathcal{L}_k\, r_{k,l}$ for the $k$th decimation step and $r=r_{0,l}$ for the full-resolution image.

Afterwards, the normalization of the statistical descriptor was carried out according to eqs.~\ref{Eq:NCF}.
The notation for each descriptor was adapted to account for the decimation process, \emph{i.e.} $F_{\beta,j,k}\left(r\right)$ denotes the normalized statistical descriptor $\beta$ for the phase $j$, obtained at the $k$th decimation step.
In Figure~\ref{Fig:Descriptors}, an example of the evolution of the normalized statistical descriptors is shown, from the full-resolution image to the last decimation step, corresponding to the images depicted in Figure~\ref{Fig:Decimation}.

\subsection{Deviation from the full-resolution image}

\begin{figure}
\centering
\includegraphics[width=0.27\textwidth]{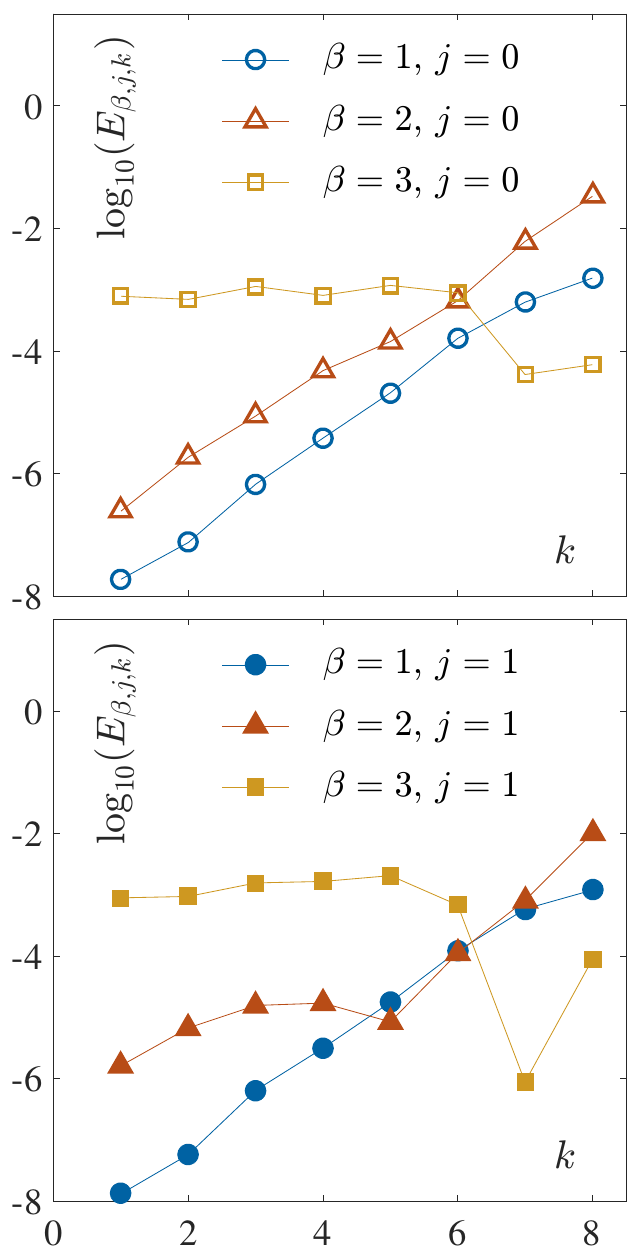}
\caption{Deviation of statistical descriptors, with respect to the full-resolution image, as a function of the $k$-th step along the random decimation process, presented in Figure~\ref{Fig:Decimation}. These deviations were obtained by applying Eq.~\eqref{Eq:Ebjk}. The subindices $\beta$ and $j$ correspond to the normalized statistical descriptor and phase, respectively, specified in section~\ref{subsec:Nsd}.}
\label{Fig:Error}
\end{figure}

As it can be discerned from the example presented in Figure~\ref{Fig:Descriptors}, the quality of the statistical descriptors reflects the information that is lost along the decimation process.
Due to the normalization of the statistical descriptors, the aforementioned quality can be separated into two components: 1) a constant shift on the surface fraction $\phi_j$ and the interfacial area per unit volume $s/\phi_j$ of the corresponding phase, and 2) a shape discrepancy with respect to the trend displayed by the statistical descriptors of the full-resolution image.
The former is directly measured from the difference between the full-resolution value and the surface fraction at the $k$th decimation step.
The latter is quantified by computing a general deviation, \emph{i.e.} the square of the difference between the statistical descriptors of $\mathcal{A}_0$ and $\mathcal{A}_k$, both at the same distance $r=\mathcal{L}_k\, r_{k,l}$, and then taking the average over all the distances.
This deviation of the descriptor $\beta$ of phase $j$ and from the image generated at the $k$th decimation step, with respect to the full-resolution image, is defined as follows
\begin{equation}
E_{\beta,j,k}=\dfrac{1}{N_k^{\dagger}}\sum_{l=1}^{N_k^{\dagger}}\Big\{ F_{\beta,j,0} \left(\mathcal{L}_k\, r_{k,l}\right)-F_{\beta,j,k}\left(\mathcal{L}_k\, r_{k,l}\right)\Big\}^2 \ .
\label{Eq:Ebjk}
\end{equation}
In Figure~\ref{Fig:Error}, computed for both phases of the OD material which decimation is displayed in Figure~\ref{Fig:Decimation}, the corresponding deviation from the full-resolution image is presented.

\begin{figure}
\centering
\includegraphics[width=0.31\textwidth]{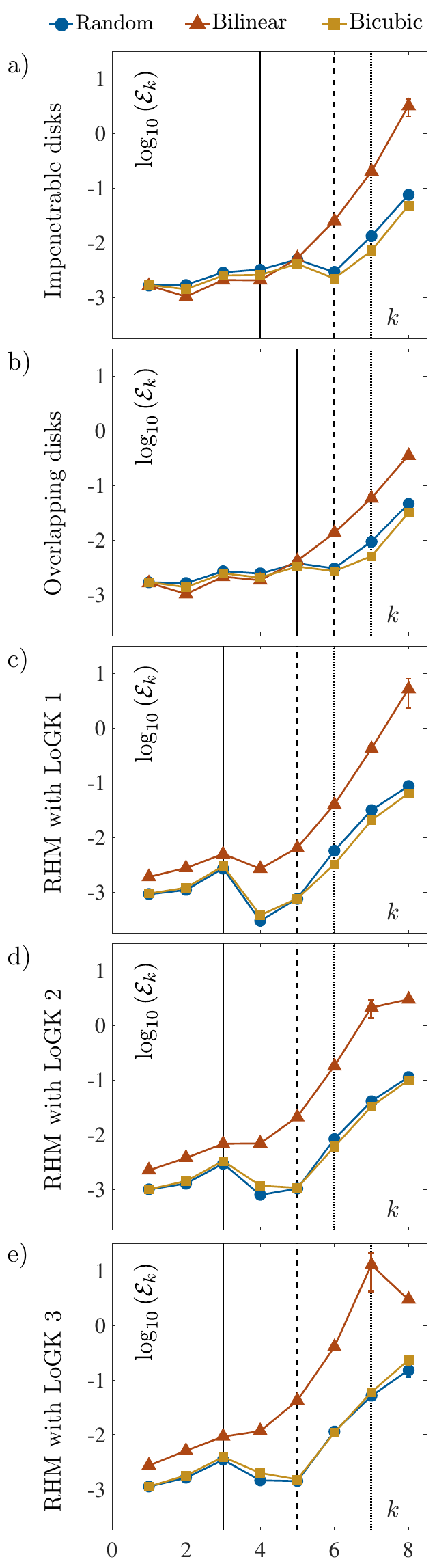}
\caption{Ensemble average global error $\langle\mathcal{E}_k\rangle$, for each material described in Tables~\ref{Tab:Disks} and \ref{Tab:CoarseGrained}, as a function of the $k$-th step along the 3 proposed decimation processes. The average and standard deviations (error bars) were obtained by applying Eqs.~\eqref{Eq:EAGErr}.
The statistically sensitive decimation step $Z$ was computed with Eqs.~\ref{Eq:DIS}, applied to the extrema, [\textbf{---}] $\min\left(\ell_{\beta,j}\right)$ and [$\boldsymbol{\cdots}$] $\max\left(\ell_{\beta,j}\right)$, and the average, [\textbf{- -}] $\overline{\ell_{\beta,j}}$, of the correlation lengths of the different descriptors $\beta$ and phases $j$.}
\label{Fig:Threshold}
\end{figure}

Moreover, the sum over the three statistical descriptors and the two phases gives rise to an overall (global) error for the $k$th decimation step 
\begin{equation}
\mathcal{E}_k=\sum_{\beta=1}^3 \sum_{j=0}^1 E_{\beta,j,k} \ .
\label{Eq:GlobalErr}
\end{equation}
This global error was calculated for each of the realizations, leading us to define an ensemble average and the corresponding standard deviation
\begin{subequations}
\begin{align}
\left\langle\mathcal{E}_k\right\rangle &=\dfrac{1}{W}\sum_{w=1}^{W} \mathcal{E}_k\left(w\right) \ , \\
\sigma_{\mathcal{E}_k} &=\sqrt{\dfrac{1}{W}\sum_{w=1}^{W} \mathcal{E}_k^2\left(w\right)-\left\langle\mathcal{E}_k\right\rangle^2}
\end{align}
\label{Eq:EAGErr}
\end{subequations}
which gathers the statistical information, over the $W=20$ configurations of each material, of the deviation observed with respect to the full-resolution images.
In Figure~\ref{Fig:Threshold}, the ensemble average is shown as a function of the decimation step, while the corresponding standard deviation appears as error bars.
The interpretation of the data presented in Figures~\ref{Fig:Decimation}--\ref{Fig:Threshold} is the central part of our analysis, thus more details and an interpretation of these results will be presented and discussed in the following section.

\subsection{Characteristic length-scale}
\label{Sec:Cls}

\begin{figure}
\centering
\includegraphics[width=0.29\textwidth]{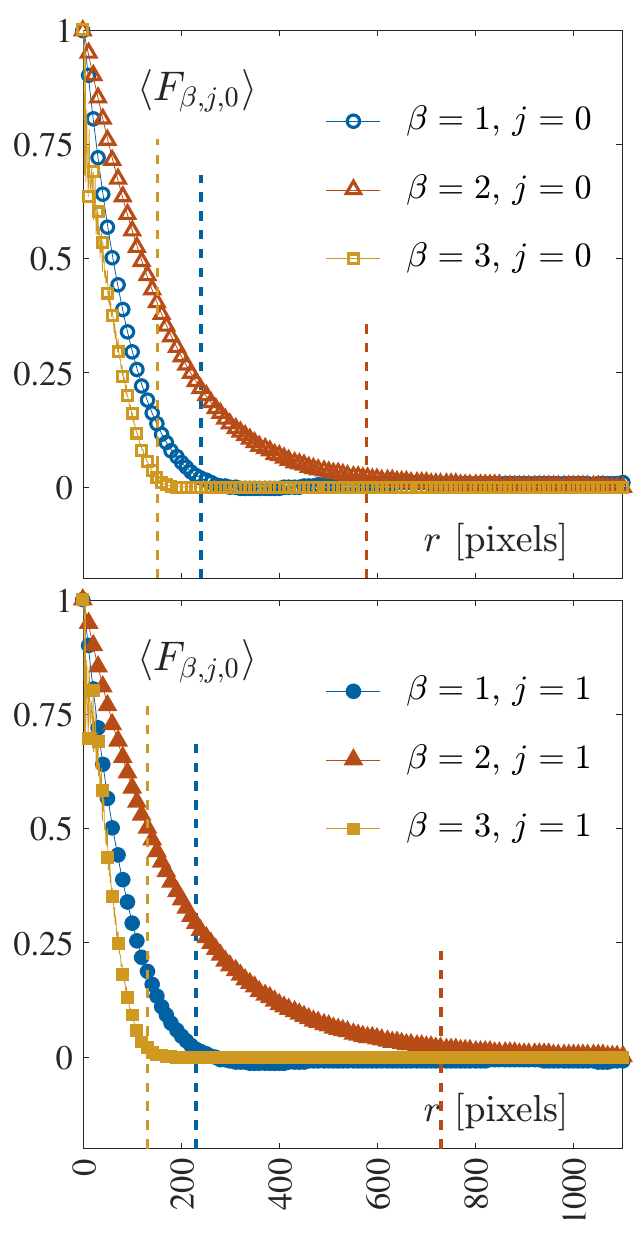}
\caption{Average statistical descriptors of the original images (with $k=0$, $M_0=N_0=4096$) of the $W=20$ different configurations of partially overlapping disks, with the same characteristics as the configuration presented in the top row, right column of Figure~\ref{Fig:Decimation}. Each average statistical descriptor was computed with Eq.~\eqref{Eq:Fbj0Av} and its corresponding correlation length is shown as a vertical dashed line. The subindices $\beta$ and $j$ correspond to the normalized statistical descriptor and phase, respectively, specified in section~\ref{subsec:Nsd}. For better visualization, only $1/10$ of the data points are shown.}
\label{Fig:Average}
\end{figure}

In addition, the ensemble average of each normalized statistical descriptor $\langle F_{\eta,j,0}\rangle$ is computed for the full-resolution images $\mathcal{A}_0$
\begin{align}
\left\langle F_{\beta,j,0}\right\rangle\left(r\right) &=\dfrac{1}{W}\sum_{w=1}^{W} F_{\beta,j,0}\left(r\right) \ ,
\label{Eq:Fbj0Av}
\end{align}
Once more, as an example, the resulting data for the partially overlapping disks is shown in Figure~\ref{Fig:Average}.

Some useful information can be extracted from these ensemble average descriptors.
For instance, one can define a correlation length $\ell_{\beta,j}$ as the range over which the descriptor $\beta$ of the phase $j$ approaches, to a certain extent, the horizontal axis $\left\langle F_{\beta,j,0}\right\rangle=0$ for the first time.
Indeed, one can define the approach extent as a margin of width 2\% of the maximum value of the function (0.02 since the correlation functions had been normalized) around the horizontal line $\left\langle F_{\beta,j,0}\right\rangle=0$.
In general, for statistically homogeneous two phases materials, $\left[0,\ell_{\beta,j}\right]$ indicates the range over which the statistical descriptor $F_{\beta,j,0}$ is non-negligible.
Consequently, one can take either the average or any order statistic (of the two phases and the three statistical descriptors) as a characteristic length-scale $\ell$ of the structural features of a given material, and the observed details on these features depend on the number of pixels that are used to depict a region of size $\ell$.
Further insight on the adequate choice of the characteristic length-scale $\ell$ will be given in the results section of this manuscript.

\subsection{Loss of statistical information}

The application of a decimation process to an image is always ensued by a loss of information.
In the case of a two-phase material configuration, the details below the size of the new pixels are lost, and this effect is reflected on the quality of the information given by the corresponding statistical descriptors.

\begin{figure}
\centering
\includegraphics[width=0.27\textwidth]{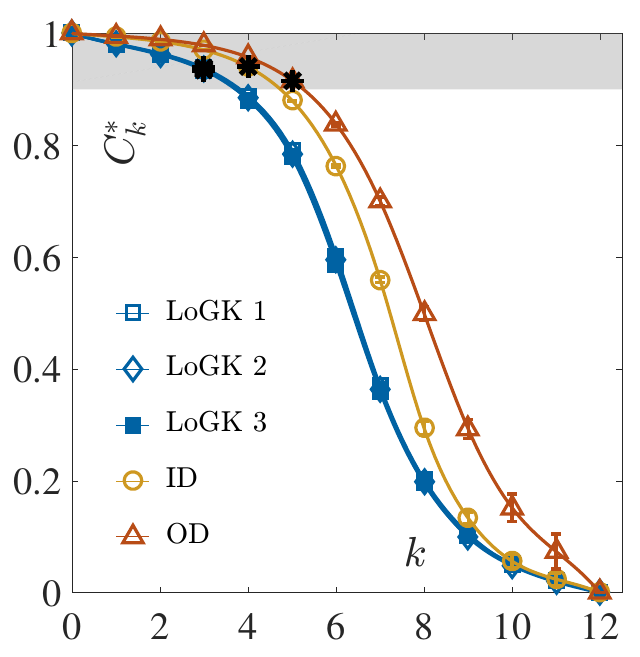}
\caption{Normalized coarseness (average and standard deviation of $W=20$ configurations) as a function of the $k$-th step along the decimation process, for each material described in Tables~\ref{Tab:Disks} and \ref{Tab:CoarseGrained}. The curves were generated with the use of Eq.~\eqref{Eq:NormCoars}. Note that the curves corresponding to the LoGK configurations overlap. The $\boldsymbol{\ast}$ symbol at each curve indicates the optimal decimation step $Z$ computed from its characteristic length-scale $\ell=\min(\ell_{\beta,j})$ and Eqs.~\eqref{Eq:DIS}. The gray surface indicates the region for which $C^{\ast}_k>0.9$.}
\label{Fig:Coarseness}
\end{figure}

Consider the variance of the local surface fraction within a square observation window with the size of a pixel at the $k$-th decimation step, which surface within the full-resolution image corresponds to an area of $2^{2k}$.
One should recall and adjust Eqs.~(\ref{Eq:ObWin}--\ref{Eq:varLVF}), for which $\bm{x}$ now indicates the center of the square observation window, with a side-length $l=2^k$.
If, at each decimation step, the normalized coarseness, given by
\begin{equation}
C^{\ast}_k=C^{\ast}\left(2^k\right)=\dfrac{\sigma\left(2^k\right)}{\sqrt{\phi_0\phi_1}} \ ,
\label{Eq:NormCoars}
\end{equation}
is computed for an observation window of the size that a pixel of the reduced-size image should cover on the full-resolution image, then the amount of information that we are losing at each decimation step is related to the complement of the normalized coarseness $1-C^{\ast}_k$.
Therefore, the closer is $C^{\ast}_k$ to one, the more information the $k$-th step decimated image retains from the original full-size image.
In contrast, if $C^{\ast}_k$ approaches zero, most of the information has been lost at the $k$-th decimation step.
In Figure~\ref{Fig:Coarseness}, the evolution of the normalized coarseness is shown as a function of the $k$-th step along the random decimation process, for all the materials generated for this study.
The behavior of $C^{\ast}_k$ is discussed in the following section.

\section{Results}
\label{Sec:Results}

A total of $W=20$ random configurations for each of the five different types of materials, which characteristics are described in Tables~\ref{Tab:Disks} and \ref{Tab:CoarseGrained}, were generated.
Each configuration is represented by a full-resolution image, to which the methodology described in the first paragraph of section~\ref{Sec:Method} was applied.
Along the first half of this section, the described results correspond specifically to the configuration of partially overlapping disks presented in Figure~\ref{Fig:Materials}, which is subjected to a random decimation process.
Nevertheless, the whole analysis is valid for all the materials and all the decimation methods.

Each time a decimation step is performed, and consequently the number of pixels decreases, some information about the configuration of the phases is lost.
This effect is clearly observed in Figure~\ref{Fig:Decimation}, where the characteristic structures of the material are distinguishable during most of the decimation process, for $k<6$.
Despite the loss of details and the apparent increase of interfacial roughness, one can still identify the $k=0$ configuration shown in the full-resolution image (top right configuration in Figure~\ref{Fig:Materials}) within the decimated image $k=5$ (Figure~\ref{Fig:Decimation}e).
A further step $k\geq6$ in the decimation process (Figure~\ref{Fig:Decimation}f-k) complicates the recognition of the original features of the material.
This qualitative analysis is supported by the quantitative results presented in Figure~\ref{Fig:Descriptors}.
For instance, as it is shown in Figure~\ref{Fig:Descriptors}a, the surface fractions $\phi_0$ and $\phi_1$ are nearly constant for $k<6$.
For $k\geq6$, $\phi_0$ increases and $\phi_1$ decreases, since $\phi_1=1-\phi_0$, both with a monotonous and moderate behavior.

In contrast, the interface between the two phases is significantly modified along the whole decimation process.
From being smooth (up to the pixel resolution) for $k=0$, as the decimation process goes on, the interface becomes sharper, as the pixel size grows.
The rounded and well defined boundaries of the phase $j=0$ presented in Figure~\ref{Fig:Materials}, become sharp and rough, more strongly for $k\geq 4$.
This is exhibited by the monotonic increase of $s/\phi_0$ and $s/\phi_1$ observed in Figure~\ref{Fig:Descriptors}a, which presents its steepest slope around $k=5$.
For the presented case, since $\phi_0\sim\phi_1$ and both being close to 0.5 despite the decimation stage, the curves for $s/\phi_0$ and $s/\phi_1$ are quite similar, but in a general situation, for which $\phi_0\nsim\phi_1$, their behavior is generally different from each other.

Additionally, when the number of pixels used to describe the material configuration decreases, the normalized statistical descriptors evolve.
For instance, the normalized autocovariance $F_{1,j,k}=\left[\chi_j^{\ast}\right]_k$ and line-path $F_{2,j,k}=\left[L_j^{\ast}\right]_k$ functions throughout the decimation process, depicted in Figure~\ref{Fig:Descriptors}, separate from the corresponding trends indicated by the correlation functions of the full-resolution image $F_{\beta,j,0}$ with $\beta=1,2$ respectively.
This discrepancy from the reference curve $F_{\beta,j,0}$ is negligible for $k\leq6$, but becomes significant and increasingly exposed as $k$ grows beyond $k=6$.
The effect of the decimation on the normalized pore-size distribution function $F_{3,j,k}=\left[P_j^{\ast}\right]_k$ is not discerned easily as compared with the two previously mentioned functions.
This is manly due to the fact that $F_{3,j,k}$ presents a noisy signal at small distances $r<150$ pixels, due to the pixelization of the images and the consequent discretization of the correlation functions.
In addition, for considerably decimated images ($k\geq6$), the gradual loss of resolution in the curves hinders the rift from the reference function $F_{3,j,0}$. 

The quantitative deviations $E_{\beta,j,k}$ of the statistical descriptor $F_{\beta,j,k}$ from the reference descriptor $F_{\beta,j,0}$, for the functions $\beta=1,2,3$ and the phases $j=0,1$, throughout the random decimation process, are computed by means of Eq.\eqref{Eq:Ebjk} and represented in Figure~\ref{Fig:Error}, for the configuration of partially overlapping disks (top-right sample in Figure~\ref{Fig:Materials} and Figure~\ref{Fig:Decimation}).
For $\beta=1$ and $\beta=2$ (corresponding to $\chi_j^{\ast}$ and $L_j^{\ast}$, respectively), the general trend of the deviations $E_{\beta,j,k}$ is an exponential growth in terms of the decimation step $k$.
In contrast, the deviation for $\beta=3$ (corresponding to $P_j^{\ast}$) is nearly constant along the first steps of the decimation process, due to its noisy behavior for $k\leq6$, whereas for $k>6$, the poor resolution of the statistical descriptor provokes a magnitude drop of the deviation $E_{3,j,k}$.

For each of the materials described in Tables~\ref{Tab:Disks} and \ref{Tab:CoarseGrained}, the full-resolution image was gradually decimated and the deviations of each descriptor $\beta$ of the phase $j$ were computed, at each $k$th decimation step.
Using Eq.~\ref{Eq:GlobalErr}, the global error was recovered, and then the evolution of the ensemble average $\langle\mathcal{E}_k\rangle$ and the standard deviation $\sigma_{\mathcal{E}_k}$, in terms of the step $k$, were obtained with Eqs.~\ref{Eq:EAGErr}.
The corresponding results are presented in Figure~\ref{Fig:Threshold}.
Despite the particular characteristics of each material and the applied decimation method, a general trend is observed: an almost constant value of $\langle\mathcal{E}_k\rangle$ along the first decimation steps, followed by a nearly exponential growth of $\langle\mathcal{E}_k\rangle$ when $k$ increases.
The standard deviation $\sigma_{\mathcal{E}_k}$ is barely perceptible in the presented semi-log graph for $k<6$, since its order of magnitude is much smaller than that of the average value.
It is also interesting to notice that, for small values of $k$ where $\mathcal{E}_k$ is almost constant, the trend followed by the global error is nearly independent of the decimation method.
Nevertheless, one realizes that the random and bicubic methods produce lower errors than the bilinear method.

In Figure~\ref{Fig:Average}, the ensemble average of the statistical descriptors are displayed, which were obtained from $W=20$ configurations represented by full-resolution images of partially overlapping disks.
The typical behavior of statistically homogeneous and isotropic materials is observed: monotonically decreasing functions with the values of $\langle F_{\beta,j,0}\rangle=1$ at $r=0$ pixels and $\langle F_{\beta,j,0}\rangle=0$ at $r\rightarrow0$ pixels, due to the corresponding normalization stage.
From each curve, the correlation length $\ell_{\beta,j}$ is determined, as stated in subsection~\ref{Sec:Cls}, and displayed in Figure~\ref{Fig:Average}.

Now, if one may accurately describe the descent of a statistical descriptors, for instance with a cubic spline, the sampling spatial rate must be at least $3/\ell_{\beta,j}$.
This means that the distance between two consecutive pixels in the decimated image should not be larger than $r=\ell_{\beta,j}/3$, a length measured in units of pixels of the full-resolution image.
Hence, the gradual decimation process should be stopped at the optimal decimation step $Z$, given by
\begin{subequations}
\begin{equation}
Z=\bigg\lfloor\min\left\{\log_2\left(\dfrac{M_0}{M_Z}\right)\ ,\log_2\left(\dfrac{N_0}{N_Z}\right)\right\}\bigg\rfloor \ ,
\end{equation}
\begin{align}
M_Z &=\bigg\lceil\dfrac{3M_0}{\ell_{\beta,j}}\bigg\rceil \ , &
N_Z &=\bigg\lceil\dfrac{3N_0}{\ell_{\beta,j}}\bigg\rceil \ ,
\end{align}
\label{Eq:DIS}
\end{subequations}
in order to provide a good statistical representation, via the statistical descriptor $F_{\beta,j,k}$, of the full-resolution image.
Briefly, $k=Z$ is the decimation step up to which one may proceed to reduce the size of the full-resolution image, preserving an acceptable amount of statistical information in the descriptor $F_{\beta,j,k}$ with respect to the reference function $F_{\beta,j,0}$.

Additionally, the extrema and the average values of $\ell_{\beta,j}$ were deduced and the corresponding decimation step $Z$, for each statistic, was computed with Eqs~.\eqref{Eq:DIS}.
The different values of $Z$ are included in Figure~\ref{Fig:Threshold}, for the materials analized in this study.
In all the cases, it is clearly observed that the step $Z$ obtained with $\min\left(\ell_{\beta,j}\right)$ corresponds to a decimation step just before the rise of the average global error $\langle\mathcal{E}_k\rangle$ occurs, whereas the ones obtained with $\overline{\ell_{\beta,j}}$ and $\max\left(\ell_{\beta,j}\right)$ are sequentially located at the already developped increasing trend of $\langle\mathcal{E}_k\rangle$. 
Thence, in order to limit the deviation from the original statistical information by 
preserving the behavior of all $F_{\beta,j,0}$, it is reasonable to chose the minimum $\ell=\min\left(\ell_{\beta,j}\right)$ as the characteristic length-scale.
In this way, $\ell$ corresponds to the maximum range over which all the normalized statistical descriptors $F_{\beta,j,k}$ are non-negligible.
Not surprisingly, in most cases, the minimum value of $\ell_{\beta,j}$ corresponds to the correlation length of the normalized pore-size distribution function $F_{3,j,0}=P_j^{\ast}$.

In Figure~\ref{Fig:Coarseness}, the coarseness $C^{\ast}_k$ is shown as a function of the $k$-th decimation step.
A reverse sigmoid function describes accurately the behavior of the curves for all the materials.
For the full-size image, the normalized coarseness is $C^{\ast}_0=1$, since the complete statistical information is available for the reference image.
As it was expected thanks to the qualitative results, most of the statistical information is preserved for the first decimation steps, which extent depends on the particular material, and the coarseness remains in the range $0.9<C^{\ast}_k<1$.
This is the case of the selected optimal number of decimation steps $Z$, obtained with $\ell=\min\left(\ell_{\beta,j}\right)$ and Eqs.~\ref{Eq:DIS}, which coarseness is $C^{\ast}_k>0.9$.
Following the decimation process, the coarseness is abruptly diminished as $k$ increases, slowing down its descent as $k$ approaches its last stage $k=12$, finally losing the statistical information when $C^{\ast}_k=0$.

\section{Algorithm}
\label{Sec:Algorithm}

From the presented results, and mainly due to the behavior of the global error $\mathcal{E}_k$ and the coarseness $C^{ \ast}$ with regards on the characteristic length-scale $\ell$, we find that $Z$ is the last decimation step before a significant rise of the global error $\mathcal{E}_k$ and an important loss of statistical information occurs.
Therefore, according to the gathered experience, a statistically sensitive decimation process is proposed, consisting in the following sequence of steps:
\begin{enumerate}
\item Compute the statistical descriptors $F_{\beta,j,0}$ of the full-resolution image $\mathbf{A}_0$.
\item Obtain the correlation lengths $\ell_{\beta,j}$ for each statistical descriptor $\beta$ and each phase $j$.
\item Determine the characteristic length-scale as $\ell=\min\left(\ell_{\beta,j}\right)$. 
\item Compute the optimal number of decimation steps $Z$ using Eqs.~\eqref{Eq:DIS}.
\item Apply $Z$ sequential steps of the selected gradual decimation process (random, bilinear or bicubic) to generate the decimated image $\mathbf{A}_Z$.
\end{enumerate}
This algorithm ensures the reduction in size of the image from $A_0\left(M_0,N_0\right)$ to $A_Z\left(M_Z,N_Z\right)$, with $M_0>M_Z$ and $N_0>N_Z$, while preserving the quality of the information stored in the statistical descriptors of the final image $\mathbf{A}_Z$ with respect to the original image $\mathbf{A}_0$.

\section{Application on SEM images}
\label{Sec:AppSEM}

To validate the presented analysis and the statistically sensitive decimation process, proposed in this study, we tested the methodology on images representing real material samples, which were obtained by SEM.

\begin{figure*}
\centering
\includegraphics[width=0.80\textwidth]{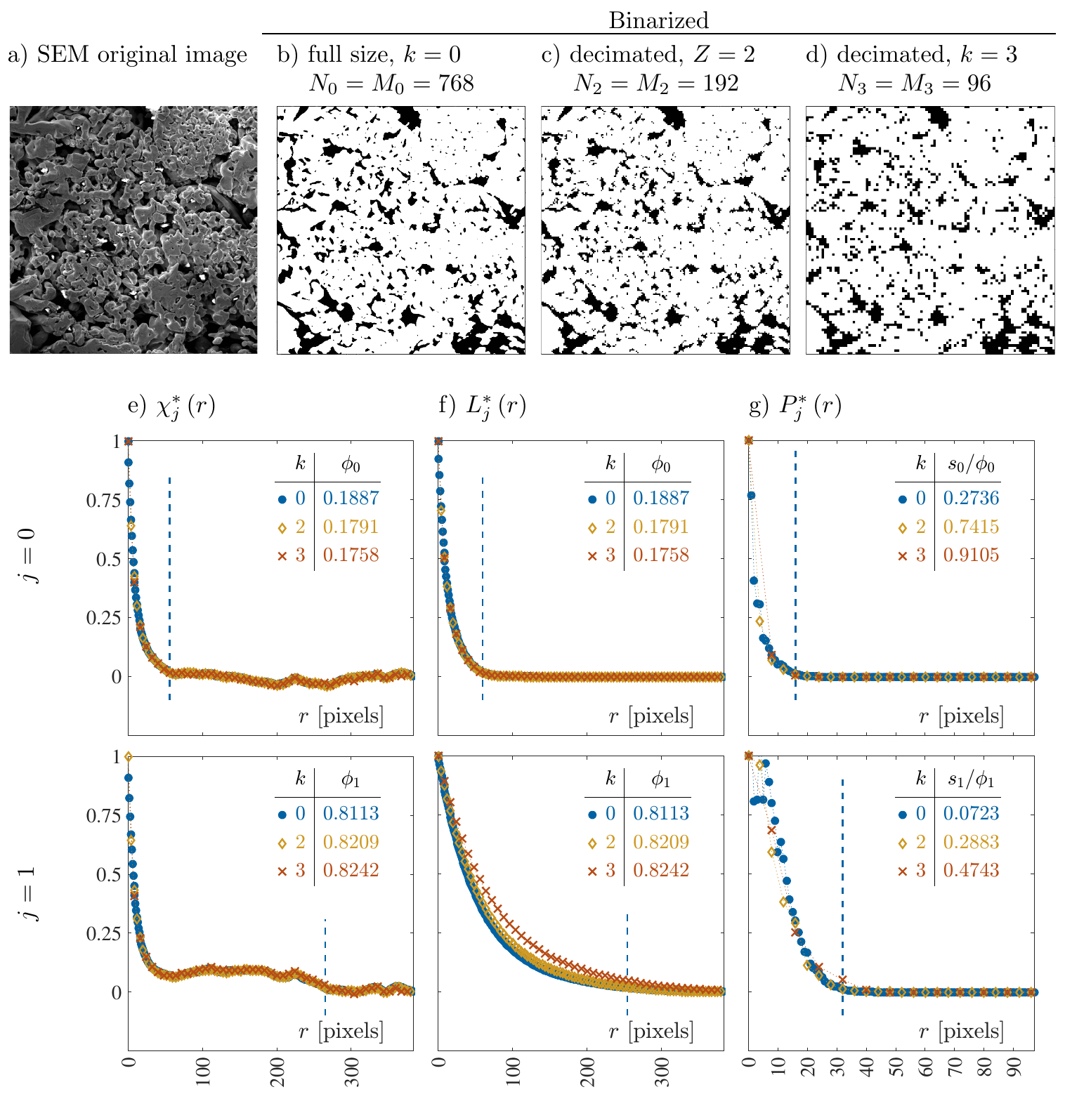}
\caption{a) Original image of a catalytic layer obtained by SEM. b) Full size binarized image. Decimated image after c) 2 steps and d) 3 steps of a bicubic process. e)-g) Statistical descriptors for the full size and the decimated images. Top and bottom rows represent the descriptors for phases $j=0$ and $j=1$ respectively, whereas each column indicates the type of descriptor: e) normalized autocovariance function $F_{1,j,k}=\chi_j$, f) normalized line-path correlation function $F_{2,j,k}=L_j$ and g) normalized pore-size distribution function $F_{3,j,k}=P_j$. In columns e)-g), filled circles (blue) correspond to the descriptors computed for the full size binarized image, whereas empty diamonds (yellow) and crosses (red) correspond to the descriptors computed for a 2-steps and 3-steps decimated binarized images, respectively.}
\label{Fig:SEMDeci}
\end{figure*}

\begin{figure*}
\centering
\includegraphics[width=0.80\textwidth]{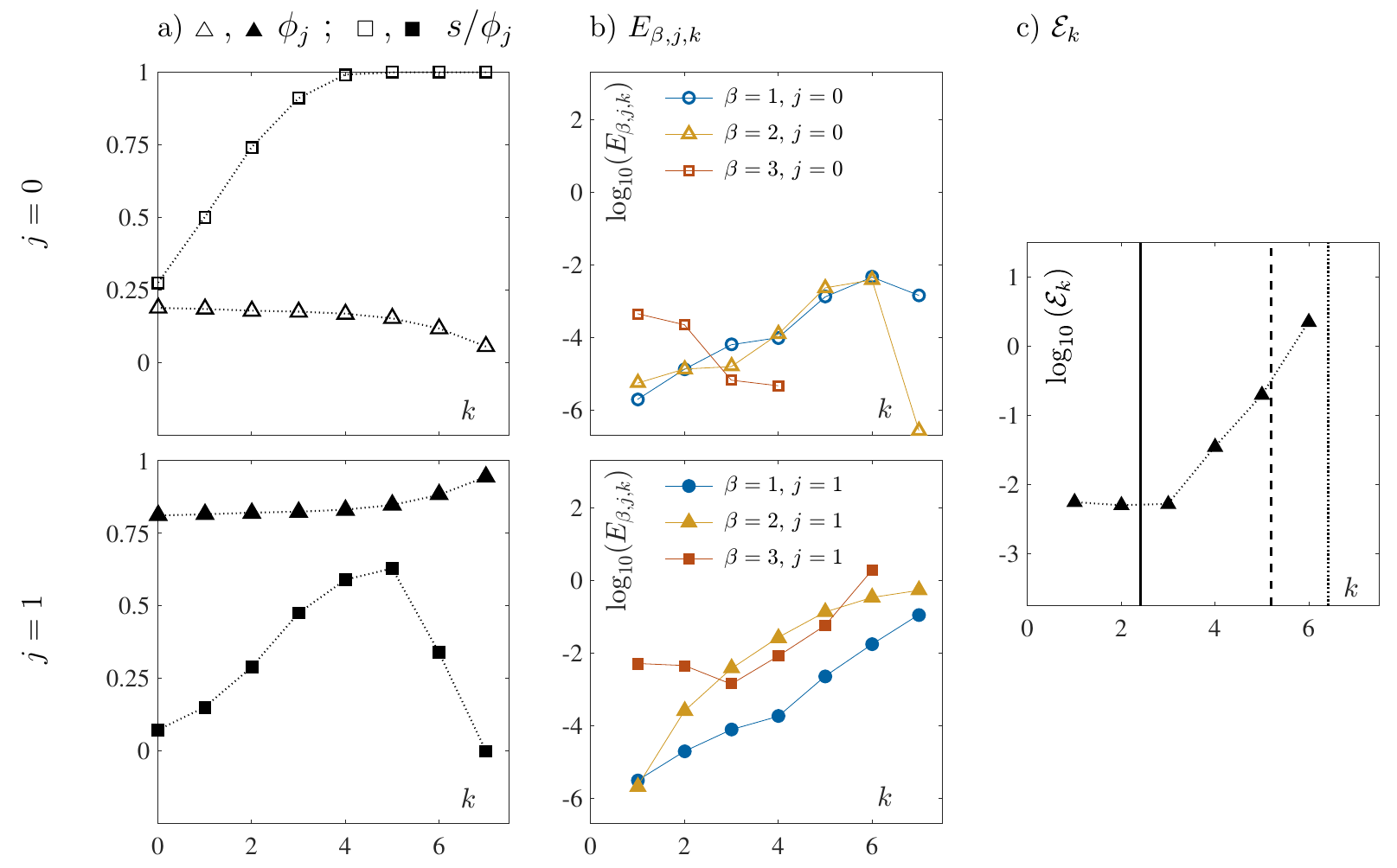}
\caption{Quantitative information about the deviation from the full-resolution SEM image presented in Figure~\ref{Fig:SEMDeci}. a) Surface fraction $\phi_j$ and specific interface area $s/\phi_j$, b) deviation of statistical descriptors $E_{\beta,j,k}$ (for each descriptor $\beta$ and phase $j$, specified in section~\ref{subsec:Nsd}.) and c) global error as functions of the $k$-th step of a bicubic decimation process.}
\label{Fig:SEMError}
\end{figure*}

A sample of the catalytic layer of a Proton Exchange Membrane Full Cell (PEMFC) was scanned and digitized with a conventional SEM, retrieving a 768$\times$768 grayscale image.
This image was binarized, following the procedure given in a previous study~\cite{OrtegonEtal2017}, yielding a binary image of the same size as the grayscale one.
Afterwards, the algorithm that has been exposed in the previous section was applied, and a decimated image with optimal size was obtained after $k=Z=2$ decimation steps.
An additional decimation step $k=3$ was performed, in order to verify that our methodology yields the desired results.
In Figure~\ref{Fig:SEMDeci}, the original SEM image, the full-size binary image and the reduced-size images after $Z=2$ and $k=3$ decimation steps are presented with their corresponding normalized statistical descriptors.
The details on the full-size image (reference) are replicated by the $Z=2$ decimated image, which is supported by the almost indistinguishable difference between each of the statistical descriptors ($\chi_j^{\ast}$, $L_j^{\ast}$ and $P_j^{\ast}$) computed from the two images, for both phases.
In contrast, despite the likeness of the $k=3$ decimated image with the full-size image, fine details had disappeared, mainly at the interface where smoothness has been lost.
This is reflected on the statistical descriptors, essentially on the line-path and the pore-size distribution functions of the $j=1$ phase, which separate substantially from their corresponding reference descriptors.
These qualitative insights are verified by the quantitative information presented in Figure~\ref{Fig:SEMError}, where the deviation of each descriptor is studied along the entire decimation process from the first step $k=1$ to a maximum and final step $k=7$.
The evolution of the surface fractions $\phi_j$ and the specific interface area per phase surface fraction $s/\phi_j$, for each phase $j$ are shown.
This values indicate the shift and dilation of the curves representing the statistical descriptors of the $k$-th decimation step, with respect to the reference descriptors.
One observes a reduction of the amount of scarce phase $\phi_0$, while for the abundant phase $\phi_1$ increases, according to the relation $\phi_0+\phi_1=1$, as the decimation process evolves.
In turn, $s/\phi_0$ grows from its initial value close to 0.25 and saturates to 1 for $k\geq4$, whereas $s\phi_1$ increases from a close to zero value towards a maximum at $k=5$ and then drops to zero at the end of the decimation process.
The variation of the deviation in shape $E_{\beta,j,k}$ of each descriptor $\beta$ and phase $j$ at the $k$-th decimation step is also reported.
The deviations of $\chi_j^{\ast}$ and $L_j^{\ast}$ show an increase as the decimation process advances, with a faster tendency for the scarce phase $j=0$ than for the abundant phase $j=1$.
$P_0^{\ast}$ shows an initial descent, but above $k=4$ it becomes negligible due to the reduced amount of data and resemblance with the reference curve.
$P_1^{\ast}$ presents a significant rise in magnitude as $k$ advances, which is comparable with the trend shown by $\chi_1^{\ast}$ and $L_1^{\ast}$.
As a consequence of the described behaviors of all the statistical descriptors, the overall error $\mathcal{E}_k$ displays a nearly constant value in the range $1\leq k\leq 3$, and a sudden exponential growth for $k>3$.
From the correlation lengths of each statistical descriptor, which are shown in Figure~\ref{Fig:SEMDeci}, the optimal number of decimation steps $Z=2$ was found.
This value $Z=2$ points at a decimation step before the sudden rise of the global error occurs, as it was expected thanks to the analysis presented in the previous sections of this manuscript.
 
The statistically sensitive decimation process exposed in this study should be applied directly to full-resolution images with numbers of rows and columns which are $M_0\propto 2^Z$ and $N_0\propto 2^Z$, respectively.
If it is not the case, an interpolation method at each decimation step must be implemented.
In any case, a gradual size reduction up to $Z$ decimation steps can be applied, being confident of the preservation of the statistical information, whenever the number of decimation steps $Z$ has been obtained by following the methodology hereinabove described.

\section{Conclusions}
\label{Sec:Conclusions}

In this manuscript, a gradual decimation process for the size reduction of images, that represent two-phases material configurations, has been developed. 
We present a methodology to define the maximum number of stages of the gradual decimation process that one can apply to an image, in order to maintain the statistical information, but to reduce its amount.
It is convenient to reduce the size of an image to perform calculations in limited computing resources, time and memory capacities. 
For instance, the reconstruction of 3D representations of random heterogeneous materials requires a memory allocation space of $N^3$ bits and approximately $N^{3x}$ operations, where $x$ represents an exponent that encloses the combined effect of the number of statistical descriptors employed to characterize the original 2D sample and the computational cost of calculating them all.
Therefore, it is essential to reduce $N$ as much as possible.
As it has been tested in this study, an unprocessed SEM image of around $800^2$ pixels will be transformed into a $800^3=5.12\times 10^8$ pixels 3D-reconstruction image.
The reduction of size after $k=2$ decimation steps, without any substantial loss of statistical information, allows us to manage a 3D-reconstruction image of only $200^3=8\times 10^6$ pixels, and to reduce the number of operations from $10^{8x}$ to $10^{6x}$.

The proposed normalization of the statistical descriptors allows us to decouple the deviation on the statistical descriptors, of gradually decimated images from the full-resolution images, in two parts: 1) a constant shift (and dilation for the two-point correlation function) on $\phi_j$ and $s/\phi_j$, and 2) a shape discrepancy, represented by the corresponding error $E_{\beta,j,k}$ for each of the correlation functions (with $\beta=1,2,3$).
Moreover, the presented analysis allows us to define a methodology to calculate an optimal number of decimation steps, based on the correlation length of the statistical descriptors, to reduce the size of an image and to minimize the time consumed for the characterization of the image, but preserving the statistical information.
Once 1) $Z$, the optimal number of decimation steps, has been defined, 2) the gradual decimation process has been performed and 3) the normalized correlation functions of the reduced-size image have been computed, one can reconstruct the statistical descriptors of the full-resolution image just by de-normalizing the reduced-size correlation functions, using the full-resolution values of $\phi_j$ and $s/\phi_j$, which are computed in a very fast and simple way.

In summary, the exhibited analysis and methodology may help to restrict the amount of information that one can afford to lose during a size reduction (decimation) process, when one aims to diminish the computational and memory cost.
The main objective is to dicrease the time consumed by a characterization or reconstruction technique, yet maintaining the statistical quality of the digitized sample image.

\bibliographystyle{apsrev} 

\begin{thebibliography}{31}
\expandafter\ifx\csname natexlab\endcsname\relax\def\natexlab#1{#1}\fi
\expandafter\ifx\csname bibnamefont\endcsname\relax
  \def\bibnamefont#1{#1}\fi
\expandafter\ifx\csname bibfnamefont\endcsname\relax
  \def\bibfnamefont#1{#1}\fi
\expandafter\ifx\csname citenamefont\endcsname\relax
  \def\citenamefont#1{#1}\fi
\expandafter\ifx\csname url\endcsname\relax
  \def\url#1{\texttt{#1}}\fi
\expandafter\ifx\csname urlprefix\endcsname\relax\def\urlprefix{URL }\fi
\providecommand{\bibinfo}[2]{#2}
\providecommand{\eprint}[2][]{\url{#2}}

\bibitem[{\citenamefont{Torquato}(2010)}]{Torquato2010}
\bibinfo{author}{\bibfnamefont{S.}~\bibnamefont{Torquato}},
  \bibinfo{journal}{Annu. Rev. Mater. Res.} \textbf{\bibinfo{volume}{40}},
  \bibinfo{pages}{101} (\bibinfo{year}{2010}).

\bibitem[{\citenamefont{Li}(2014)}]{Li2014}
\bibinfo{author}{\bibfnamefont{D.}~\bibnamefont{Li}}, \bibinfo{journal}{JOM}
  \textbf{\bibinfo{volume}{66}}, \bibinfo{pages}{444} (\bibinfo{year}{2014}).

\bibitem[{\citenamefont{Yeong and
  Torquato}(1998{\natexlab{a}})}]{Torquato1998a}
\bibinfo{author}{\bibfnamefont{C.}~\bibnamefont{Yeong}} \bibnamefont{and}
  \bibinfo{author}{\bibfnamefont{S.}~\bibnamefont{Torquato}},
  \bibinfo{journal}{Phys. Rev. E} \textbf{\bibinfo{volume}{57}},
  \bibinfo{pages}{495} (\bibinfo{year}{1998}{\natexlab{a}}).

\bibitem[{\citenamefont{Yeong and
  Torquato}(1998{\natexlab{b}})}]{Torquato1998b}
\bibinfo{author}{\bibfnamefont{C.}~\bibnamefont{Yeong}} \bibnamefont{and}
  \bibinfo{author}{\bibfnamefont{S.}~\bibnamefont{Torquato}},
  \bibinfo{journal}{Phys. Rev. E} \textbf{\bibinfo{volume}{58}},
  \bibinfo{pages}{224} (\bibinfo{year}{1998}{\natexlab{b}}).

\bibitem[{\citenamefont{Cule and Torquato}(1999)}]{Torquato1999}
\bibinfo{author}{\bibfnamefont{D.}~\bibnamefont{Cule}} \bibnamefont{and}
  \bibinfo{author}{\bibfnamefont{S.}~\bibnamefont{Torquato}},
  \bibinfo{journal}{J. Appl. Phys.} \textbf{\bibinfo{volume}{86}},
  \bibinfo{pages}{3428} (\bibinfo{year}{1999}).

\bibitem[{\citenamefont{Jiao et~al.}(2007)\citenamefont{Jiao, Stillinger, and
  Torquato}}]{Torquato2007}
\bibinfo{author}{\bibfnamefont{Y.}~\bibnamefont{Jiao}},
  \bibinfo{author}{\bibfnamefont{F.}~\bibnamefont{Stillinger}},
  \bibnamefont{and} \bibinfo{author}{\bibfnamefont{S.}~\bibnamefont{Torquato}},
  \bibinfo{journal}{Phys. Rev. E} \textbf{\bibinfo{volume}{76}},
  \bibinfo{pages}{031110} (\bibinfo{year}{2007}).

\bibitem[{\citenamefont{Garmestani et~al.}(2009)\citenamefont{Garmestani,
  Baniassadi, Fathi, and Ahzi}}]{GarmestaniEtal2009}
\bibinfo{author}{\bibfnamefont{H.}~\bibnamefont{Garmestani}},
  \bibinfo{author}{\bibfnamefont{M.}~\bibnamefont{Baniassadi}},
  \bibinfo{author}{\bibfnamefont{M.}~\bibnamefont{Fathi}}, \bibnamefont{and}
  \bibinfo{author}{\bibfnamefont{S.}~\bibnamefont{Ahzi}},
  \bibinfo{journal}{Int. J. of Theor. Appl. Mult. Mech.}
  \textbf{\bibinfo{volume}{1}}, \bibinfo{pages}{134} (\bibinfo{year}{2009}).

\bibitem[{\citenamefont{Barbosa et~al.}(2011)\citenamefont{Barbosa, Andaverde,
  Escobar, and Cano}}]{BarbosaEtal2011}
\bibinfo{author}{\bibfnamefont{R.}~\bibnamefont{Barbosa}},
  \bibinfo{author}{\bibfnamefont{J.}~\bibnamefont{Andaverde}},
  \bibinfo{author}{\bibfnamefont{B.}~\bibnamefont{Escobar}}, \bibnamefont{and}
  \bibinfo{author}{\bibfnamefont{U.}~\bibnamefont{Cano}}, \bibinfo{journal}{J.
  Power Sources} \textbf{\bibinfo{volume}{196}}, \bibinfo{pages}{1248}
  (\bibinfo{year}{2011}).

\bibitem[{\citenamefont{Baniassadi et~al.}(2012)\citenamefont{Baniassadi,
  Mortazavi, Amani~Hamedani, Garmestani, Ahzi, Fathi-Torbaghan, Ruch, and
  Khaleel}}]{BaniassadiEtal2012}
\bibinfo{author}{\bibfnamefont{M.}~\bibnamefont{Baniassadi}},
  \bibinfo{author}{\bibfnamefont{B.}~\bibnamefont{Mortazavi}},
  \bibinfo{author}{\bibfnamefont{H.}~\bibnamefont{Amani~Hamedani}},
  \bibinfo{author}{\bibfnamefont{H.}~\bibnamefont{Garmestani}},
  \bibinfo{author}{\bibfnamefont{S.}~\bibnamefont{Ahzi}},
  \bibinfo{author}{\bibfnamefont{M.}~\bibnamefont{Fathi-Torbaghan}},
  \bibinfo{author}{\bibfnamefont{D.}~\bibnamefont{Ruch}}, \bibnamefont{and}
  \bibinfo{author}{\bibfnamefont{M.}~\bibnamefont{Khaleel}},
  \bibinfo{journal}{Comput. Mater. Sci.} \textbf{\bibinfo{volume}{51}},
  \bibinfo{pages}{372} (\bibinfo{year}{2012}).

\bibitem[{\citenamefont{Huang and Li}(2013)}]{HuangLi2013}
\bibinfo{author}{\bibfnamefont{M.}~\bibnamefont{Huang}} \bibnamefont{and}
  \bibinfo{author}{\bibfnamefont{Y.}~\bibnamefont{Li}},
  \bibinfo{journal}{Comput. Mater. Sci.} \textbf{\bibinfo{volume}{67}},
  \bibinfo{pages}{63} (\bibinfo{year}{2013}).

\bibitem[{\citenamefont{Tahmasebi and Sahimi}(2012)}]{TahmasebiEtal2012}
\bibinfo{author}{\bibfnamefont{P.}~\bibnamefont{Tahmasebi}} \bibnamefont{and}
  \bibinfo{author}{\bibfnamefont{M.}~\bibnamefont{Sahimi}},
  \bibinfo{journal}{Phys. Rev. E} \textbf{\bibinfo{volume}{85}},
  \bibinfo{pages}{066709} (\bibinfo{year}{2012}).

\bibitem[{\citenamefont{Tahmasebi and Sahimi}(2013)}]{TahmasebiEtal2013}
\bibinfo{author}{\bibfnamefont{P.}~\bibnamefont{Tahmasebi}} \bibnamefont{and}
  \bibinfo{author}{\bibfnamefont{M.}~\bibnamefont{Sahimi}},
  \bibinfo{journal}{Phys. Rev. Lett.} \textbf{\bibinfo{volume}{110}},
  \bibinfo{pages}{078002} (\bibinfo{year}{2013}).

\bibitem[{\citenamefont{Tahmasebi and Sahimi}(2015)}]{TahmasebiEtal2015}
\bibinfo{author}{\bibfnamefont{P.}~\bibnamefont{Tahmasebi}} \bibnamefont{and}
  \bibinfo{author}{\bibfnamefont{M.}~\bibnamefont{Sahimi}},
  \bibinfo{journal}{Phys. Rev. E} \textbf{\bibinfo{volume}{91}},
  \bibinfo{pages}{032401} (\bibinfo{year}{2015}).

\bibitem[{\citenamefont{Liu et~al.}(2013)\citenamefont{Liu, Greene, Chen,
  Dikin, and Liu}}]{LiuEtal2013}
\bibinfo{author}{\bibfnamefont{Y.}~\bibnamefont{Liu}},
  \bibinfo{author}{\bibfnamefont{M.}~\bibnamefont{Greene}},
  \bibinfo{author}{\bibfnamefont{W.}~\bibnamefont{Chen}},
  \bibinfo{author}{\bibfnamefont{D.}~\bibnamefont{Dikin}}, \bibnamefont{and}
  \bibinfo{author}{\bibfnamefont{W.}~\bibnamefont{Liu}},
  \bibinfo{journal}{Computer-Aided Design} \textbf{\bibinfo{volume}{45}},
  \bibinfo{pages}{65} (\bibinfo{year}{2013}).

\bibitem[{\citenamefont{Pant et~al.}(2014)\citenamefont{Pant, Mitra, and
  M.}}]{PantEtal2014}
\bibinfo{author}{\bibfnamefont{L.}~\bibnamefont{Pant}},
  \bibinfo{author}{\bibfnamefont{S.}~\bibnamefont{Mitra}}, \bibnamefont{and}
  \bibinfo{author}{\bibfnamefont{S.}~\bibnamefont{M.}}, \bibinfo{journal}{Phys.
  Rev. E} \textbf{\bibinfo{volume}{90}}, \bibinfo{pages}{023306}
  (\bibinfo{year}{2014}).

\bibitem[{\citenamefont{Torquato}(2002)}]{Torquato}
\bibinfo{author}{\bibfnamefont{D.}~\bibnamefont{Torquato}},
  \emph{\bibinfo{title}{Random Heterogeneous Materials microscopic and
  macroscopic properties}} (\bibinfo{publisher}{Springer},
  \bibinfo{year}{2002}).

\bibitem[{\citenamefont{Sahimi}(2003)}]{Sahimi}
\bibinfo{author}{\bibfnamefont{M.}~\bibnamefont{Sahimi}},
  \emph{\bibinfo{title}{Heterogeneous Materials I: Linear transport and optical
  properties}} (\bibinfo{publisher}{Springer}, \bibinfo{year}{2003}).

\bibitem[{\citenamefont{Snarskii et~al.}(2016)\citenamefont{Snarskii,
  Bezsudnov, Sevryukov, Morozovskiy, and Malinsky}}]{Snarskii}
\bibinfo{author}{\bibfnamefont{A.}~\bibnamefont{Snarskii}},
  \bibinfo{author}{\bibfnamefont{I.}~\bibnamefont{Bezsudnov}},
  \bibinfo{author}{\bibfnamefont{V.}~\bibnamefont{Sevryukov}},
  \bibinfo{author}{\bibfnamefont{A.}~\bibnamefont{Morozovskiy}},
  \bibnamefont{and} \bibinfo{author}{\bibfnamefont{J.}~\bibnamefont{Malinsky}},
  \emph{\bibinfo{title}{Transport processes in macroscopically disordered
  media}} (\bibinfo{publisher}{Springer}, \bibinfo{year}{2016}).

\bibitem[{\citenamefont{Kanit et~al.}(2006)\citenamefont{Kanit, NGuyen, Forest,
  Jeulin, Reed, and Singleton}}]{KanitEtal2006}
\bibinfo{author}{\bibfnamefont{T.}~\bibnamefont{Kanit}},
  \bibinfo{author}{\bibfnamefont{F.}~\bibnamefont{NGuyen}},
  \bibinfo{author}{\bibfnamefont{S.}~\bibnamefont{Forest}},
  \bibinfo{author}{\bibfnamefont{D.}~\bibnamefont{Jeulin}},
  \bibinfo{author}{\bibfnamefont{M.}~\bibnamefont{Reed}}, \bibnamefont{and}
  \bibinfo{author}{\bibfnamefont{S.}~\bibnamefont{Singleton}},
  \bibinfo{journal}{Comput. Methods Appl. Mech. Engrg.}
  \textbf{\bibinfo{volume}{195}}, \bibinfo{pages}{3960} (\bibinfo{year}{2006}).

\bibitem[{\citenamefont{Sinclair}(2015)}]{Textiles2015}
\bibinfo{author}{\bibfnamefont{R.}~\bibnamefont{Sinclair}},
  \emph{\bibinfo{title}{Textiles and Fashion - Materials, Design and
  Technology}} (\bibinfo{publisher}{Elsevier}, \bibinfo{year}{2015}),
  \bibinfo{edition}{woodhead publishing series in textiles: number 126} ed.

\bibitem[{\citenamefont{Suzuki et~al.}(2010)\citenamefont{Suzuki, Hattori,
  Miura, Tsuboi, Hatakeyama, Takaba, Williams, and
  Miyamoto}}]{MiyamotoEtal2010}
\bibinfo{author}{\bibfnamefont{A.}~\bibnamefont{Suzuki}},
  \bibinfo{author}{\bibfnamefont{T.}~\bibnamefont{Hattori}},
  \bibinfo{author}{\bibfnamefont{R.}~\bibnamefont{Miura}},
  \bibinfo{author}{\bibfnamefont{H.}~\bibnamefont{Tsuboi}},
  \bibinfo{author}{\bibfnamefont{N.}~\bibnamefont{Hatakeyama}},
  \bibinfo{author}{\bibfnamefont{H.}~\bibnamefont{Takaba}},
  \bibinfo{author}{\bibfnamefont{M.}~\bibnamefont{Williams}}, \bibnamefont{and}
  \bibinfo{author}{\bibfnamefont{A.}~\bibnamefont{Miyamoto}},
  \bibinfo{journal}{Int. J. Electrochem. Sci.} \textbf{\bibinfo{volume}{5}},
  \bibinfo{pages}{1948} (\bibinfo{year}{2010}).

\bibitem[{\citenamefont{Shojaeefard et~al.}(2016)\citenamefont{Shojaeefard,
  Molaeimanesh, Nazemian, and Moqaddari}}]{ShojaeefardEtal2016}
\bibinfo{author}{\bibfnamefont{M.}~\bibnamefont{Shojaeefard}},
  \bibinfo{author}{\bibfnamefont{G.}~\bibnamefont{Molaeimanesh}},
  \bibinfo{author}{\bibfnamefont{M.}~\bibnamefont{Nazemian}}, \bibnamefont{and}
  \bibinfo{author}{\bibfnamefont{M.}~\bibnamefont{Moqaddari}},
  \bibinfo{journal}{Int. J. Hydrogen Energy} \textbf{\bibinfo{volume}{179}},
  \bibinfo{pages}{1} (\bibinfo{year}{2016}).

\bibitem[{\citenamefont{Che and Vedrine}(2012)}]{CheVedrine}
\bibinfo{author}{\bibfnamefont{M.}~\bibnamefont{Che}} \bibnamefont{and}
  \bibinfo{author}{\bibfnamefont{J.}~\bibnamefont{Vedrine}},
  \emph{\bibinfo{title}{Characterization of Solid Materials and Heterogeneous
  Catalysts: From Structure to Surface Reactivity, Volume 1\&2}}
  (\bibinfo{publisher}{Wiley-VCH Verlag GmbH \& Co. KGaA},
  \bibinfo{year}{2012}).

\bibitem[{\citenamefont{Harris and Chiu}(2015{\natexlab{a}})}]{HarrisChiuA2015}
\bibinfo{author}{\bibfnamefont{W.}~\bibnamefont{Harris}} \bibnamefont{and}
  \bibinfo{author}{\bibfnamefont{W.}~\bibnamefont{Chiu}}, \bibinfo{journal}{J.
  Power Sources} \textbf{\bibinfo{volume}{282}}, \bibinfo{pages}{552}
  (\bibinfo{year}{2015}{\natexlab{a}}).

\bibitem[{\citenamefont{Harris and Chiu}(2015{\natexlab{b}})}]{HarrisChiuB2015}
\bibinfo{author}{\bibfnamefont{W.}~\bibnamefont{Harris}} \bibnamefont{and}
  \bibinfo{author}{\bibfnamefont{W.}~\bibnamefont{Chiu}}, \bibinfo{journal}{J.
  Power Sources} \textbf{\bibinfo{volume}{282}}, \bibinfo{pages}{622}
  (\bibinfo{year}{2015}{\natexlab{b}}).

\bibitem[{\citenamefont{C.Pelissou et~al.}(2009)\citenamefont{C.Pelissou,
  Baccou, Monerie, and Perales}}]{PelissouEtal2009}
\bibinfo{author}{\bibfnamefont{C.}~\bibnamefont{C.Pelissou}},
  \bibinfo{author}{\bibfnamefont{J.}~\bibnamefont{Baccou}},
  \bibinfo{author}{\bibfnamefont{Y.}~\bibnamefont{Monerie}}, \bibnamefont{and}
  \bibinfo{author}{\bibfnamefont{F.}~\bibnamefont{Perales}},
  \bibinfo{journal}{Int. J. Solids Struct.} \textbf{\bibinfo{volume}{46}},
  \bibinfo{pages}{2842} (\bibinfo{year}{2009}).

\bibitem[{\citenamefont{Mirkhalaf et~al.}(2016)\citenamefont{Mirkhalaf,
  Andrade~Pires, and Simoes}}]{MirkhalafEtal2016}
\bibinfo{author}{\bibfnamefont{S.}~\bibnamefont{Mirkhalaf}},
  \bibinfo{author}{\bibfnamefont{F.}~\bibnamefont{Andrade~Pires}},
  \bibnamefont{and} \bibinfo{author}{\bibfnamefont{R.}~\bibnamefont{Simoes}},
  \bibinfo{journal}{Finite Elem. Anal. Des.} \textbf{\bibinfo{volume}{119}},
  \bibinfo{pages}{30} (\bibinfo{year}{2016}).

\bibitem[{\citenamefont{Zachary and Torquato}(2009)}]{Torquato2009}
\bibinfo{author}{\bibfnamefont{C.~E.} \bibnamefont{Zachary}} \bibnamefont{and}
  \bibinfo{author}{\bibfnamefont{S.}~\bibnamefont{Torquato}},
  \bibinfo{journal}{J. Stat. Mech.} \textbf{\bibinfo{volume}{12}},
  \bibinfo{pages}{P12015} (\bibinfo{year}{2009}).

\bibitem[{\citenamefont{Bovik}(2000)}]{Bovik}
\bibinfo{author}{\bibfnamefont{A.}~\bibnamefont{Bovik}},
  \emph{\bibinfo{title}{Handbook of image and video processing}}
  (\bibinfo{publisher}{Academic Press}, \bibinfo{year}{2000}).

\bibitem[{\citenamefont{Gonzalez and Woods}(2008)}]{Gonzalez}
\bibinfo{author}{\bibfnamefont{R.}~\bibnamefont{Gonzalez}} \bibnamefont{and}
  \bibinfo{author}{\bibfnamefont{R.}~\bibnamefont{Woods}},
  \emph{\bibinfo{title}{Digital image processing}}
  (\bibinfo{publisher}{Pearson-Prentice Hall}, \bibinfo{year}{2008}),
  \bibinfo{edition}{3rd} ed.

\bibitem[{\citenamefont{Ortegon et~al.}(2017)\citenamefont{Ortegon,
  Ledesma-Alonso, Barbosa, Vazquez~Castillo, and
  Castillo~Atoche}}]{OrtegonEtal2017}
\bibinfo{author}{\bibfnamefont{J.}~\bibnamefont{Ortegon}},
  \bibinfo{author}{\bibfnamefont{R.}~\bibnamefont{Ledesma-Alonso}},
  \bibinfo{author}{\bibfnamefont{R.}~\bibnamefont{Barbosa}},
  \bibinfo{author}{\bibfnamefont{J.}~\bibnamefont{Vazquez~Castillo}},
  \bibnamefont{and}
  \bibinfo{author}{\bibfnamefont{A.}~\bibnamefont{Castillo~Atoche}},
  \bibinfo{journal}{Submitted to Computational Materials Science}
  (\bibinfo{year}{2017}).

\end{thebibliography}

\end{document}